\documentclass[10pt,letterpaper]{article}
\usepackage[top=0.85in,left=2.75in,footskip=0.75in]{geometry}

\usepackage{amsmath,amssymb}

\usepackage{changepage}

\usepackage[utf8x]{inputenc}

\usepackage{textcomp,marvosym}

\usepackage{cite}

\usepackage{nameref,hyperref}

\usepackage[right]{lineno}

\usepackage{microtype}
\DisableLigatures[f]{encoding = *, family = * }

\usepackage[table]{xcolor}

\usepackage{array}

\newcolumntype{+}{!{\vrule width 2pt}}

\newlength\savedwidth

\raggedright
\usepackage[aboveskip=1pt,labelfont=bf,labelsep=period,justification=raggedright,singlelinecheck=off]{caption}

\makeatletter
\renewcommand{\@biblabel}[1]{\quad#1.}
\makeatother

\usepackage{multirow,multicol}

\usepackage{lastpage,fancyhdr,graphicx}
\usepackage{epstopdf}
\fancyheadoffset[L]{2.25in}
\fancyfootoffset[L]{2.25in}
\begin{document}
\vspace*{0.2in}

\begin{flushleft}
{\Large
\textbf\newline{Regional economic integration via detection of circular flow in international value-added network} 
}
\newline
\\
Sotaro Sada*,
Yuichi Ikeda\textsuperscript{\ddag}
\\
\bigskip
Graduate School of Advanced Integrated Studies in Human Survivability, Kyoto University, Kyoto, Japan
\\
\bigskip

%
%





* sada.sotaro.87a@st.kyoto-u.ac.jp \\
\ddag~ikeda.yuichi.2w@kyoto-u.ac.jp

\end{flushleft}
\section*{Abstract}

Global value chains (GVCs) are formed through value-added trade, and some regions promote economic integration by concluding regional trade agreements to promote these chains. However, there is no way to quantitatively assess the scope and extent of economic integration involving various sectors in multiple countries.
In this study, we used the World Input--Output Database to create a cross-border sector-wise trade in value-added network (international value-added network (IVAN)) covering the period of 2000--2014 and evaluated them using network science methods.
By applying Infomap to the IVAN, we confirmed for the first time the existence of two regional communities: Europe and the Pacific Rim.
Helmholtz--Hodge decomposition was used to decompose the value flows within the region into potential and circular flows, and the annual evolution of the potential and circular relationships between countries and sectors was clarified. The circular flow component of the decomposition was used to define an economic integration index, and findings confirmed that the degree of economic integration in Europe declined sharply after the economic crisis in 2009 to a level lower than that in the Pacific Rim. The European economic integration index recovered in 2011 but again fell below that of the Pacific Rim in 2013.
Moreover, sectoral analysis showed that the economic integration index captured the effect of Russian mineral resources, free movement of labor in Europe, and international division of labor in the Pacific Rim, especially in GVCs for the manufacture of motor vehicles and high-tech products.


\section*{Introduction}

It is difficult to grow a country's economy without establishing economic relations with other countries. Every year, the amount of international trade increases along with the world's GDP. Free trade agreements (FTAs) and regional trade agreements (RTAs) have been created to support this trend, and countries are working to stabilize trade.

How, then, do countries become interdependent through trade? Classically, Ricard advocated comparative advantage, which stated that countries specializing in different industries are supposed to trade by taking advantage of their respective strengths. However, in recent years, trade in intermediate goods, which did not exist at that time, has begun and flourished. In other words, there are forms of trade specializing in different industries and GVCs in the same industry to produce complex and sophisticated products. In addition, distance (as represented by the gravity model) and economic integration (such as that promoted by the EU and NAFTA) facilitate international trade.

Many studies have been conducted on measuring international production structures. In the 21st century, measurement of GVCs became a widely discussed topic after Hummels et al. \cite{hummels} proposed the vertical specialization index. Moreover, the VAX and FVAiX indices were proposed by Johnson and Noguera \cite{johnson12} and Amador et al. \cite{ama15}, respectively. Koopman et al. \cite{koop} decomposed and classified the trade prices into nine terms and clarified double-counted terms.

The abovementioned progress has led to the development of international input--output (IO) tables. After the World Input--Output Database (WIOD) was publicized by Dietzenbacher et al. \cite{diet} and Timmer et al. \cite{timmer15} \cite{timmer}, Cerina et al. \cite{cerina} and Zhu et al. \cite{zhu} analyzed the data for measuring GVCs network analysis methods. Los et al. \cite{los16} showed that GVCs have been fragmenting more rapidly than regionally in 1995--2011. According to Johnson and Noguera \cite{johnson17}, over a period of 40 years, the ratio of value-added to gross exports in non-manufacturing sectors rose, while that in manufacturing sectors fell by 20 percentage points; their study showed that RTAs led to such decline.

Due to international relationship problems in the economy, researchers in international organizations have been analyzing GVCs. Recently comprehensive summary reports of GVC research have been published by the World Bank \cite{gvcdr17} \cite{gvcdr19}. According to these reports, GVC research aims to address misunderstanding of trade data and give international trade opportunities and export diversity to developing countries for economic growth \cite{dollar}.
As shown by Degain et al., GVCs had been expanding until the global financial crisis, but it stopped growing in 2011 when it recovered from a decline in equipment. The GVC expansion seen before the financial crisis has been interpreted as China's accession to the WTO and participation in GVCs \cite{dollar}. These GVC research developments were described in detail by Amador and Cabral \cite{ama16} and Inomata \cite{inomata}. However, economics studies mainly focus on the relationship between nations. There are limited works showing the complicated relationship between industries in different countries.

By contrast, there are many studies on international trade in network science because of recent developments in analytical methods and data availability. Initially, the scale-free characteristics of international trade networks were clarified \cite{li}\cite{serrano}\cite{fagio08}; then, the virtual nodes in these networks were examined using centrality indices and other measures, and changes over time were analyzed \cite{fagio09}\cite{foti}\cite{xiang}.

Ikeda et al. \cite{ikeda14} \cite{ikeda16} studied the international trade network of the WIOD using community analysis of modularity maximization, which combines the synchronization aspect of the G7 production network \cite{ikeda15} \cite{Aoyama}; then, they found stable sectoral communities that had emerged in 1995--2011.

From an institutional point of view, a study examining interregional agreements and geographic factors in international trade change found no evidence that the WTO has contributed to the increase in international trade and indicated that geographic factors are more significant than interregional agreements \cite{bari}. By contrast, other studies show that trade policies contribute to shaping trade networks \cite{deb}, and the impact of distance by trade is decreasing in industrial sectors \cite{borchert}.

However, there is no way to quantitatively assess the scope and extent of economic integration involving various sectors in multiple countries.
It is difficult to quantitatively evaluate such economic phenomena, where multiple countries and sectors are intertwined, and the trade balance between two countries is still used as a basis for foreign economic policy. In this study, we analyze the economic linkages through trade that have been made in recent years using international IO tables and network science methods and clarify how the world economy is being integrated from the perspective of value-added trade.

In this context, this study offers the following essential contributions. First, rather than using trade figures, which can mislead international economic relations, we used the value-added calculations used in an IO analysis to construct a sector-wise network.
Next, from the value-added linkage, we confirmed the regional characteristics of Europe and the Pacific Rim. Findings showed that the behavior of this value-added relationship is highly different from that of the sectoral community in the international trade network shown in previous studies. The dense economic relationships within these regions were decomposed into potential and circular relationships, and the countries and sectors that contribute to the strong regional ties were identified. Finally, we propose an economic integration index that is based on the circular flow component, and the results showed that economic integration was higher in the Pacific Rim than in Europe in 2010, 2013, and 2014.

The remainder of this paper is organized into three parts. The next section describes the IO table and the computational method used in this work to create the network and the community analysis, explains the decomposition of the potential and circular flow components, and presents the proposed method for measuring economic integration. Then, the characteristics and structure of the network, the potential and circular relationships, and the results of the integration index are presented, and the international economic linkages are examined on the basis of these results. Finally, we conclude this study.

\section*{Data and methods}
This section consists of four subsections. First, we describe the WIOD and the computational methods used to build the cross-border, sector-wise trade in value-adde network (IVAN). Second, we briefly describe Infomap, a community analysis algorithm, which was used to determine the extent to which economic integration is occurring. Third, we explain the use of Helmholtz--Hodge decomposition to extract the potential and circular nature of IVANs. Finally, the economic integration index is defined.

\subsection*{Data}
We used the WIOD released in 2016, which includes 43 countries and classifies the rest of the world (RoW) into 56 sectors. These countries and sectors are listed in \nameref{S1_Table} and \nameref{S2_Table}. The country codes are the same as those of the WIOD, but we simplified the sectoral codes.
The WIOD has IO tables for 2000--2014. Table~\ref{table1} shows a simplified example of the table. The IO table was developed by Leontief~\cite{leo} and is now used globally to estimate economic and environmental conditions. In this paper, we used the WIOD without RoW because we aim to analyze the relationships of specific countries.

\begin{table}[!ht]
\begin{adjustwidth}{-2.25in}{0in} 
	\begin{center}
	\caption{{\bf World input--output table with two sectors in three countries.}} \label{table1}
	\scalebox{0.8}[0.8]{
	\begin{tabular}{|c|c|c|c|c|c|c|c|c|c|c|c|} \hline
	\multicolumn{2}{|c|}{} & \multicolumn{6}{c|}{Intermediate demand} & \multicolumn{3}{c|}{\multirow{2}{*}{Final demand}} & \\ \cline{3-8}
	\multicolumn{2}{|c|}{} & \multicolumn{2}{c|}{Country A} & \multicolumn{2}{c|}{Country B} &\multicolumn{2}{c|}{Country C} & \multicolumn{3}{c|}{} & Total output \\ \cline{3-11}
	\multicolumn{2}{|c|}{} & Sector 1 & Sector 2 & Sector 1 & Sector 2  & Sector 1 & Sector 2 & Country A & Country B  & Country C &  \\ \hline
	\multirow{2}{*}{Country A}& Sector 1 & $Z_{1,1}$ & $Z_{1,2}$ & $Z_{1,3}$ & $Z_{1,4}$ & $Z_{1,5}$ & $Z_{1,6}$ & $FD_{1,1}$ & $FD_{1,2}$ & $FD_{1,3}$ & $T_{1}$ \\ \cline{2-12}
			& Sector 2 & $Z_{2,1}$ & $Z_{2,2}$ & $Z_{2,3}$ & $Z_{2,4}$ & $Z_{2,5}$ & $Z_{2,6}$ & $FD_{2,1}$ & $FD_{2,2}$ & $FD_{2,3}$ & $T_{2}$ \\ \hline
	\multirow{2}{*}{Country B}	& Sector 1 & $Z_{3,1}$ & $Z_{3,2}$ & $Z_{3,3}$ & $Z_{3,4}$ & $Z_{3,5}$ & $Z_{3,6}$ & $FD_{3,1}$ & $FD_{3,2}$ & $FD_{3,3}$ & $T_{3}$ \\ \cline{2-12}
			& Sector 2 & $Z_{4,1}$ & $Z_{4,2}$ & $Z_{4,3}$ & $Z_{4,4}$ & $Z_{4,5}$ & $Z_{4,6}$ & $FD_{4,1}$ & $FD_{4,2}$ & $FD_{4,3}$ & $T_{4}$ \\ \hline
	\multirow{2}{*}{Country C}& Sector 1 & $Z_{5,1}$ & $Z_{5,2}$ & $Z_{5,3}$ & $Z_{5,4}$ & $Z_{5,5}$ & $Z_{5,6}$ & $FD_{5,1}$ & $FD_{5,2}$ & $FD_{5,3}$ & $T_{5}$\\ \cline{2-12}
				& Sector 2 & $Z_{6,1}$ & $Z_{6,2}$ & $Z_{6,3}$ & $Z_{6,4}$ & $Z_{6,5}$ & $Z_{6,6}$ & $FD_{6,1}$ & $FD_{6,2}$ & $FD_{6,3}$ & $T_{6}$ \\ \hline
	 \multicolumn{2}{|c|}{Value-added} & $VA_{1}$ & $VA_{2}$ & $VA_{3}$ & $VA_{4}$ & $VA_{5}$ & $VA_{6}$ & & & & \\ \hline
	  \multicolumn{2}{|c|}{Total output} & $T_{1}$ & $T_{2}$ & $T_{3}$ & $T_{4}$ & $T_{5}$ & $T_{6}$ & & & & \\ \hline
	 \end{tabular}
	 }
	 \end{center}
\end{adjustwidth}
\end{table}

\subsection*{Calculation for adjacency matrix of IVAN}
Let $n_c$ be the number of countries and $n_s$ be the number of sectors in the IO table. Then, we define the $n_c n_s \times n_c n_s$ intermediate matrix $\mathbf{Z}$, $n_c n_s \times n_c$ final demand matrix $\mathbf{FD}$, $1 \times n_c n_s$ value-added vector $\mathbf{VA}$, and $1 \times n_c n_s$ total output matrix $\mathbf{T}$.
With use of the value-added vector $\mathbf{VA}$ and total output vector $\mathbf{T}$, the value-added coefficient vector $\mathbf{V}$ is calculated as $\mathbf{VA}/(\mathbf{T})^\top$.
Therefore, an induced value-added vector $\hat{\mathbf{V}}\mathbf{L}\mathbf{F}$ is made by a vector $\mathbf{L}\mathbf{F}$ left-hand multiplied by the $n_c n_s \times n_c n_s$ diagonal matrix $\hat{\mathbf{V}}$, where the diagonal components are the vector $\mathbf{V}$.

Then, the adjacency matrix of the global value-added network (GVAN) $\mathbf{G}$ is calculated as $\hat{\mathbf{V}}\mathbf{L}\hat{\mathbf{F}}$, where $\hat{\mathbf{F}}$ is a diagonal matrix whose diagonal components are the vector $\mathbf{F}$. The IVAN's adjacency matrix $\mathbf{Y}$ was created by eliminating components of 43 on-diagonal $56\times56$ blocks as zero from $\mathbf{G}$. This matrix represents the sum of all the ripple effects of value-added induced by the final demand in foreign sectors. The GVAN and IVAN nodes comprise 56 sectors in 43 countries. \nameref{S1_Table} and \nameref{S2_Table} list these countries and sectors.

In summary, IVANs are directed and weighted networks constructed by the adjacency matrix $Y$, which does not have domestic links. By contrast, GVANs have domestic and international links.

\subsection*{Community detection}
To know how each network develop a community structure, we apply community analysis to IVANs. Some community analysis methods have been developed, and their features were summarized by Fortunato and Hric \cite{fortu16} and Barab\'{a}si \cite{barabasi}.
We used Infomap, which is an application of random walk and Huffman coding \cite{huffman} to analyze the communities in the studied network \cite{mapsof08}. Let a random walker run in the network where this analysis will be applied; the random walker transitions with a probability dependent on the path weights (plus a constant transposition probability) \cite{rankncluster}. The map equation establishes communities and renames the node in which the random walk track's code length is minimized.
In comparison, the well-known method of modularity maximization clusters a network by counting nodes' link weights, inflows, and outflows, whereas the map equation method clusters a network by the remaining time of the random walker in the nodes; this difference leads to variations in results \cite{mapsof08} \cite{rosvall}.

In the Infomap method, the optimization problem for community segmentation in a network is replaced with the minimization problem of the code length of a segmented network.
Consider the segmentation of a network composed of $n$ nodes into $m$ communities with a community partition $\mathsf{M}$. Let the mean code length of segmented communities (index code length) be $H(\mathcal{Q})$ and the mean code length of nodes within a community $i$ be $H(\mathcal{P}^i)$; then, with use of Shannon's source coding theorem \cite{shannon}, the average description length of a single step of the random walk $L(M)$ is calculated as follows:
\begin{equation} \label{mapequation}
L(\mathsf{M}) = q_\curvearrowright H(\mathcal{Q}) + \sum_{i=1}^m p_\circlearrowright^i H(\mathcal{P}^i),
\end{equation}
where $q_\curvearrowright$ is the probability of the random walker moving to another community and $p_\circlearrowright^i$ is the probability of the random walker moveing within a community $i = 1,2,\ldots,m$ plus the exit probability from $i$. Each of the probabilities is defined as $q_\curvearrowright = \sum_{i=1}^m q_{i\curvearrowright}$ and $p_\circlearrowright^i = \sum_{\alpha\in i}p_\alpha + q_{i\curvearrowright}$, where $p_\alpha$ is the probability of visiting node $\alpha = 1,2,\ldots,n$ and $q_{i\curvearrowright}$ is the exit probability from community $i$. Eq~(\ref{mapequation}) is called the map equation \cite{rosvall}.

\subsection*{Structural characteristics of networks}
Certain indices reveal network characteristics. We used eleven indices: (1) density, reciprocity, diameter, average path length, average betweenness, assortativity, average in-degree, and average out-degree are calculated as unweighted and directed networks; (2) in-strength and out-strength are for weighted and directed ones; (3) the clustering coefficient is calculated as unweighted and undirected \cite{barabasi}.

The density of a network is the proportion of the number of links to the number of maximum possible links in the network. Thus, it is calculated as $l/n(n-1)$, where $n$ and $l$ are the numbers of nodes and links in the network, respectively.

The reciprocity of a network is the proportion of the reciprocal links to the total number of links in the network. It is calculated as $l_m/l$, where $l_m$ is the number of reciprocal links (two directed links between two nodes in the network).

The clustering coefficient indicates how each node is connected to its neighbors. The clustering coefficient $c_i$ of a node $i$ is calculated as $c_i=l_i/k_i(k_i-1)$, where $l_i$ ($k_i$) is the number of links (neighbors) of node $i$. Thus, the clustering coefficient of a network is $\sum c_i/n$.

The diameter of a network is the maximum number of shortest paths for all pairs of nodes in the network. 

The average path length of a network is the mean of the shortest path length in the network: $\sum_{i,j; i\neq j} d_{ij}/n(n-1)$, where $d_{ij}$ is the shortest path length between nodes $i$ and $j$.

The betweenness is calculated by the number of the shortest paths through a node. The betweenness $b_i$ of node $i$ is $\sum_{j\neq k\neq i}s_{jk}(i)/s_{jk}$, where $s_{jk}$ is the number of the shortest paths between nodes $j$ and $k$, and $s_{jk}(i)$ is the number of the shortest paths through node $i$. Therefore, the average betweenness of a network is the mean of $b_i$.

The degree of a node means the sum of the number of links in the node. There are two kinds of degrees in a directed network, namely, in-degree and out-degree, which are the sums of the links to the node from the other nodes (inflow) and links from the node to other nodes (outflow), respectively. Thus, the in-degree and out-degree are represented as: $\sum_{j} l_{j,i}$ and $\sum_{j} l_{i,j}$, respectively, where $l_{i,j}$ is a link (flow) from node $i$ to node $j$.

The assortativity of a network scores the similarity of the connections in the network as -1 to 1, which is calculated by $(\sum_{jk}jke_{jk}-\mu)/\sigma^2$, where $e_{jk}$ is the probability of the two nodes of degrees $j$ and $k$ are at the end of a randomly chosen link, and $\mu$ and $\sigma$ are the mean and standard deviation, respectively, of the excess degree distribution. Assortativity that is close to -1 (1) indicates that high-degree nodes in a network are connected to low-degree nodes (high-degree nodes).

The strength of a node is the total weight of the links in the node. There are two kinds of strength, in a directed network, namely, in-strength and out-strength, which are the summed weights of the inflow and outflow, respectively. Thus, the in-strength and out-strength are represented as: $\sum_{j} l^w_{j,i}$ and $\sum_{j} l^w_{i,j}$, respectively, where $l^w_{i,j}$ is the flow amount from node $i$ to node $j$.

\subsection*{Decomposition to potential and circular flows}
Firstly, we define flows as directed and weighted links, and the flow of IVANs as value flow.
We used Helmholtz--Hodge decomposition~\cite{hhd} to extract potential and circular relationships from IVANs. This method was also used by Kichikawa et al.~\cite{kichi} to clarify the potential relationships in corporate transaction networks.
With Helmholtz--Hodge decomposition, the flow $F_{ij}$ from node $i$ to node $j$ can be separated into the circular flow $F_{ij}^{(c)}$ and potential flow $F_{ij}^{(p)}$: $ F_{ij} = F_{ij}^{(c)} + F_{ij}^{(g)} $.
Here the potential flow $ F^{(p)}_{ij} $ is given by $ F^{(p)}_{ij} =w_{ij}(\phi_i -\phi_j)$, where $\phi_i$ is the Helmholtz--Hodge potential and $w_{ij}$ is a positive weight for the link between node $i$ and node $j$.
The circular flow satisfies $\sum_j F^{(c)}_{ij}=0$, in which the inflows and outflows are balanced in each node.
$F_{ij}^{(p)}$ represents the difference in potentials between the nodes, whereas $F_{ij}^{(c)}$ represents the number of feedback loops among the nodes.

An IVAN captures the value-added chain of various international sectors to their respective final goods. The potential of a sector is the difference between the amount of value adde by the sector for final foreign production and the amount of value added by the foreign sector for the sector's final production. In other words, if the value-added potential is positive, the sector contributes more to the production of the final goods of the foreign sector, and if it is negative, the sector is contributed more to the own production of the final goods by foreign sectors. This can indicate the degree of asymmetric dependence or co-dependence. Moreover, the circular flow component indicates the amount of contribution to (from) the foreign sector, that is, the degree of interdependence.

\subsection*{Economic integration index}
When several economies proceed with economic integration by some factors such as the establishment of FTAs or RTAs, the value added in a country becomes increasingly induced by other countries, and vice versa. In other words, higher economic integration makes a larger feedback loop within an IVAN. On the basis of this assumption, we define the economic integration index $E$ as the aggregate amount of circular flow divided by the total flow of the GVAN in a specific community.
\begin{equation} \label{eii}
E = \frac{\sum_{i>j} Y_{ij}^{(c)}}{\sum G_{ij}}
\end{equation}
The range of application of this index covers communities detected by Infomap, which also detects communities using flow. The detected communities are interpreted as the circulation observed when random walkers rotate in the community. The degree of circulation is quantified and divided by the economic scale of the community; then, the economic integration is measured as the amount of value-added circulation per economic scale.

The index defined here indicates how much of the domestic and international value-added chain toward final demand can be extracted as international interdependence. The higher the value of the index, the greater the value-added induced in the community across national borders (economic activities performed by sectors in other countries), and economic integration in this study is measured by this value.

This economic integration index can decompose into sectoral indices arbitrarily. For sector $k$, the sectoral economic integration index is
\begin{equation} \label{eii}
E_k = \frac{\sum_{i>j; i,j\in S_k} Y_{ij}^{(c)}}{\sum_{i,j\in S_k} G_{ij}},
\end{equation}
where $S_k$ is a set of nodes in sector $k$.

In summary, the economic integration index is the circular magnitude of the cross-border value-added contribution of economic activities in a country (other countries) toward the final production of sectors in other countries (home country), compared with the circular magnitude of the value-added contribution of economic activities in the community, including circulation within a country.

\section*{Results} 
This section is composed of three subsections. The first subsection shows the community structure of the IVANs, the threshold set for community detection, and the resulting 15-year change in regional communities. The second subsection shows the basic characteristics of the original IVAN, the cut IVAN, and the IVAN of the regional communities in selected years. The third subsection confirms the value-added potential and circular relationships revealed by Helmholtz--Hodge decomposition for countries and sectors in Europe and the Pacific Rim. The last subsection shows the evolution of economic integration in the regional communities and the results of decomposing the economic integration indicators into key sectoral components.

\subsection*{Communities of IVANs}
\begin{figure}[!h]
\begin{adjustwidth}{-2.25in}{0in} 
\includegraphics[width=18cm,keepaspectratio]{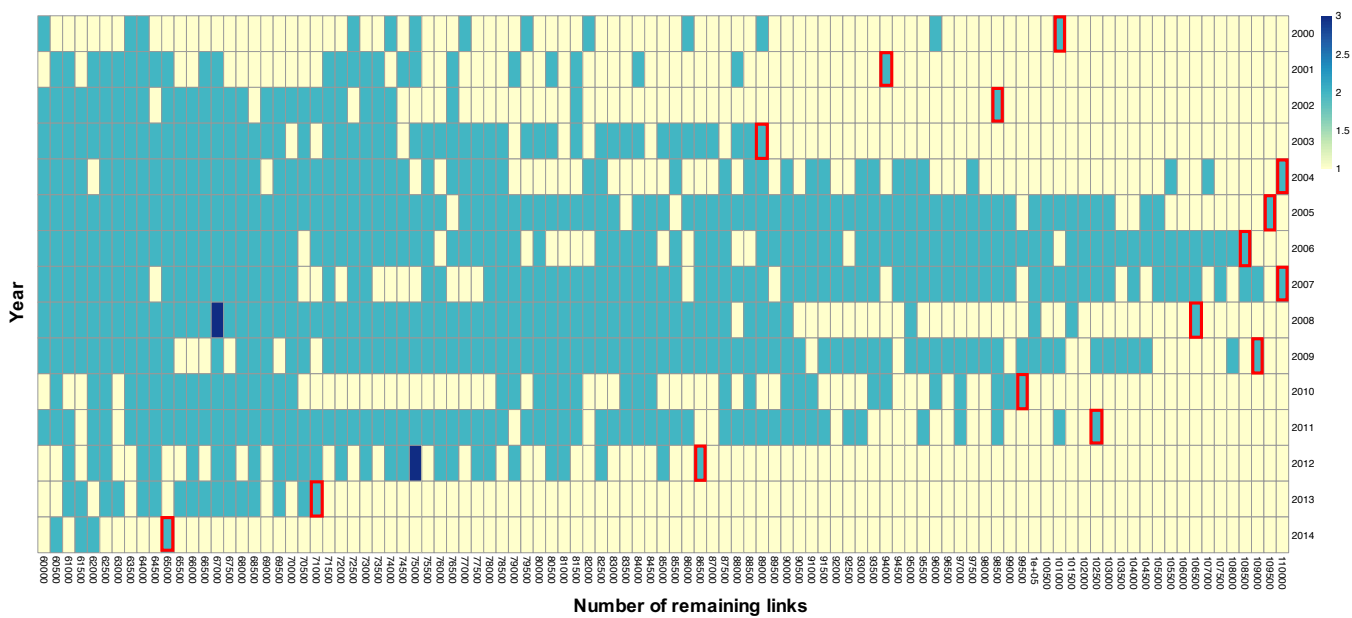}
\caption{{\bf Heat map of the number of large communities in the IVANs cut by each threshold, in 2000--2014.}
This research uses the thresholds represented as a red cell in each year. Colors means the number of communities.}
\label{fig1}
\end{adjustwidth}
\end{figure}

Direct application of Infomap to the IVANs reveals one large community for 15 years. This result differs from the community of international trade network shown in a previous study.
Next, thresholds are set to exclude the branches and leaves of the IVANs to see any concealed community structure. In the range of 6,500--11,000 remaining links, several large communities appear (Fig~\ref{fig1}). Here, the thresholds are set by the number of the most extensive links to be retained, as the threshold value set by USD is inappropriate for the weight of the IVAN links, which fluctuates due to the economic growth and economic crisis that occurred in those 15 years. In Fig~\ref{fig1}, each cell colored on the number of communities with more than 240 nodes which is 10\% of the total nodes in the IVANs. The remaining number of links in 2014 is almost half of that in 2004--2009.

\begin{figure}[!h]
\begin{adjustwidth}{-2.25in}{0in} 
\centering
\includegraphics[width=18cm,keepaspectratio]{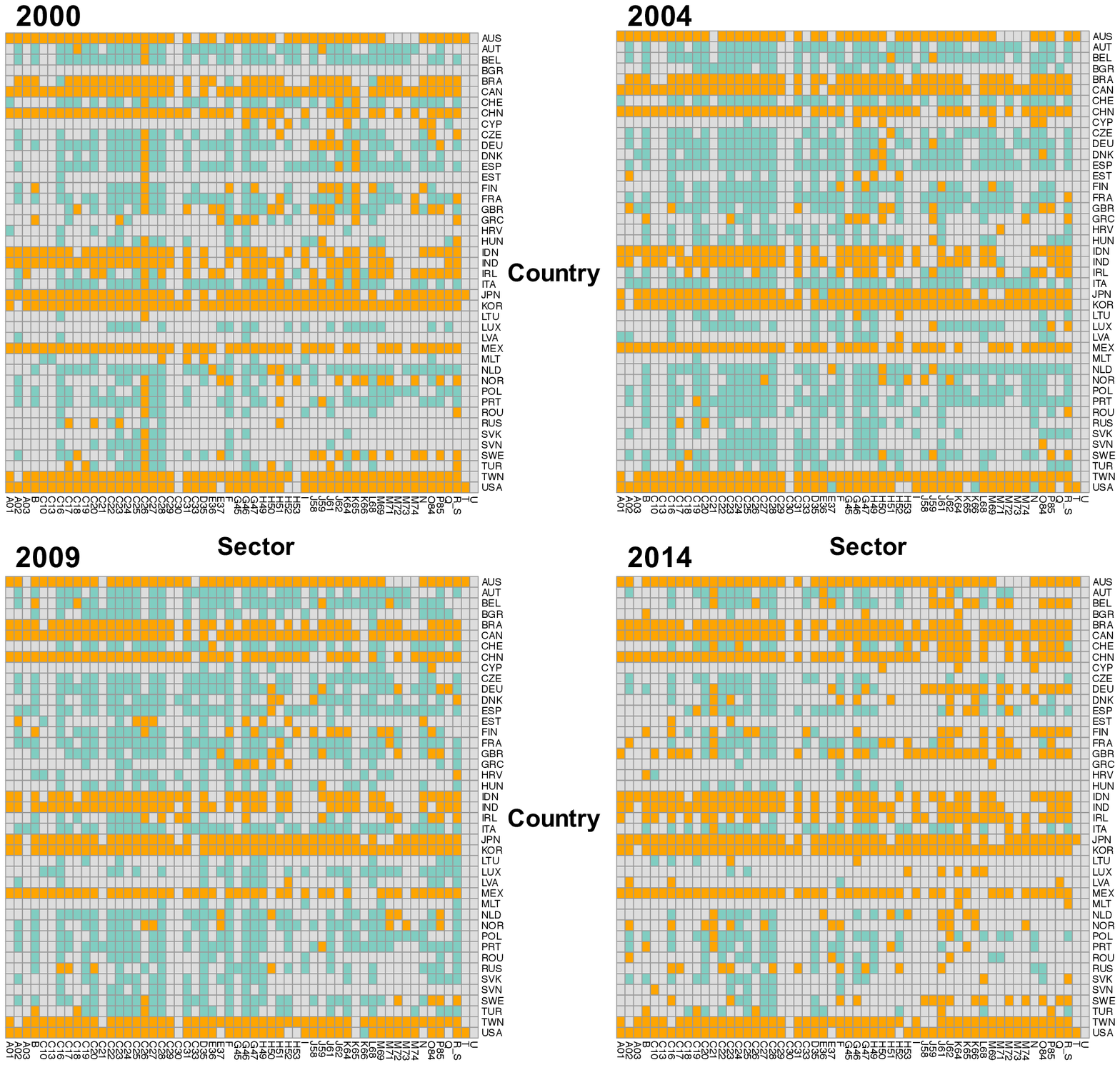}
\caption{{\bf Community heat maps in 2000, 2004, 2009, and 2014.}
Each cell is a node of IVAN, the gray nodes are not a communities, while the other colors indicate community. The green and orange cells are nodes of the European and Pacific Rim communities, respectively. The regional classification is in \nameref{S1_Table}.}
\label{fig2}
\end{adjustwidth}
\end{figure}

By using the abovementioned threshold values, we can detect the communities from the IVANs as shown in Fig~\ref{fig2} which shows four years' heat maps and represents each cell as a node of the IVANs.
In the maps, the separated communities are indicated by different colors; the gray cells are the nodes that were not considered communities. The figure shows orange horizontal stripes, which include most of the sectors in countries such as Australia, Brazil, and Canada. The green nodes include the sectors A01: agriculture, C29: manufacture of motor vehicles, and C30: manufacture of other transport equipment of mainly European countries. The gray area includes many other sectors in Europe, especially in small countries.

\begin{figure}[!h]
\begin{adjustwidth}{-2.25in}{0in} 
\centering
\includegraphics[width=18cm,keepaspectratio]{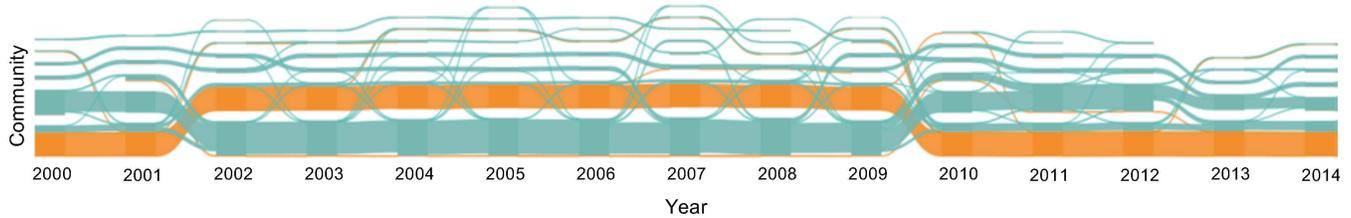}
\caption{{\bf Sankey diagram of communities in 2000--2014.}
Communities are ordered by size, from the largest to the smallest, each year. Green and orange represent nodes of Europe and the Pacific Rim, respectively. Their classification of them is in \nameref{S1_Table}. The two bottom communities are used to denote regional communities, but there have been other small European communities over the years.}
\label{fig3}
\end{adjustwidth}
\end{figure}

The results of community detection 15 years are shown in Fig~\ref{fig3} as a Sankey diagram, where orange represents the Pacific Rim sectors, and the green represents the European sectors (nodes) (the details of the classification are in \nameref{S1_Table}), and we can see that these two regions are detected as separate communities throughout the 15 years.

\subsection*{Observed structural characteristics}
Table~\ref{table2} shows the structural features for 2000, 2008, and 2014. The densities and cluster coefficients show that the IVANs are dense, and the reciprocity shows that more than 97\% of the links are mutual.
The density of the IVANs that are cut at the threshold in Fig.~\ref{fig1} is reduced by about 82\%. In terms of reciprocities, the cut IVANs are about 20\% of the remaining links that are cut. This indicates that the IVAN have many asymmetric mutualities with large values in only one direction. In the two local communities, the cluster coefficients and reciprocities are at the same level as those of the IVANs, although the densities are about 10\% higher than those of the entire IVANs.
\begin{table}[!ht]
\begin{adjustwidth}{-2.25in}{0in} 
	\centering
	\caption{{\bf Structural characteristics of IVANs, IVANs cut with the threshold, IVANs in the regional communities: Europe and the Pacific Rim.}} \label{table2}
	\scalebox{0.75}[0.75]{
	\begin{tabular}{c|cccc|cccc|cccc} \hline
	 & \multicolumn{4}{c|}{2000} & \multicolumn{4}{c|}{2008} & \multicolumn{4}{c}{2014} \\ \cline{2-13}
	 & IVAN & Cut IVAN & Europe & Pacific Rim & IVAN & Cut IVAN & Europe & Pacific Rim & IVAN & Cut IVAN & Europe & Pacific Rim \\ \hline
	Density & 0.831 & 0.0174 & 0.942 & 0.921 & 0.841 & 0.0184 & 0.955 & 0.925 & 0.831 & 0.0112 & 0.945 & 0.929 \\ \hline
	Reciprocity & 0.972 & 0.231 & 0.987 & 0.976 & 0.982 & 0.192 & 0.990 & 0.987 & 0.974 & 0.169 & 0.993 & 0.975 \\ \hline
	Clustering coefficient & 0.976 & 0.321 & 0.953 & 0.941 & 0.976 & 0.319 & 0.963 & 0.934 & 0.976 & 0.278 & 0.950 & 0.950 \\ \hline
	Diameter & 2 & 5 & 2 & 2 & 2 & 4 & 2 & 2 & 2 & 6 & 2 & 2 \\ \hline
	Average path length & 1.023 & 2.474 & 1.046 & 1.056 & 1.023 & 2.55 & 1.035 & 1.063 & 1.023 & 2.656 & 1.048 & 1.048 \\ \hline
	Average betweenness* & 51.70 & 2644 & 24.62 & 37.14 & 51.72 & 3053 & 28.99 & 37.37 & 51.64 & 2835 & 14.32 & 35.82 \\ \hline
	Assortativity & -0.0167 & -0.436 & -0.0394 & -0.0304 & -0.0179 & -0.427 & -0.0294 & -0.0305 & -0.0170 & -0.424 & -0.0505 & -0.0304 \\ \hline
	Average in-degree* & 2196 & 71.68 & 516.4 & 622 & 2198 & 71 & 788 & 556.8 & 2193 & 52.59 & 281.7 & 715.3 \\ \hline
	Average out-degree* & 2143 & 68.71 & 509.7 & 610.7 & 2164 & 65.5 & 780.3 & 553 & 2147 & 45.61 & 279.8 & 703 \\ \hline
	 \end{tabular}
	 }
	\begin{flushleft} *These averages were calculated for nonzero values.
	\end{flushleft}
\end{adjustwidth}
\end{table}

Each IVAN has a diameter of 2, which means that any sector in any country can be connected to all sectors through one sector in another country.
The minimum diameter is 2 because links to sectors in the same country are not included, but since the average path length is close to 1, most sectors are directly connected.
In the cut IVANs, the diameter is about 5, and the average betweenness is more than 50 times larger than that of the original IVANs because the diameter and average path length are long, and a particular node mediates many shortest paths. In the regional communities, the average path length is a slightly longer, and the betweenness is roughly half of the original IVANs.

The number of nodes in the IVANs is 2,408, but the average in-degree and out-degree value is over 2,100. Since most of the sectors are connected, the assortativity is close to zero.
In the cut IVANs, the degrees and assortativity are around 70 and -0.43, respectively, indicating that sectors with degree differences are connected by high values flows.
In the regional communities, the degree varies according to the size of the detected regional community, and the assortativity is negative, close to zero.

\begin{figure}[!h]
\begin{adjustwidth}{-2.25in}{0in}
\centering
\includegraphics[width=15cm,keepaspectratio]{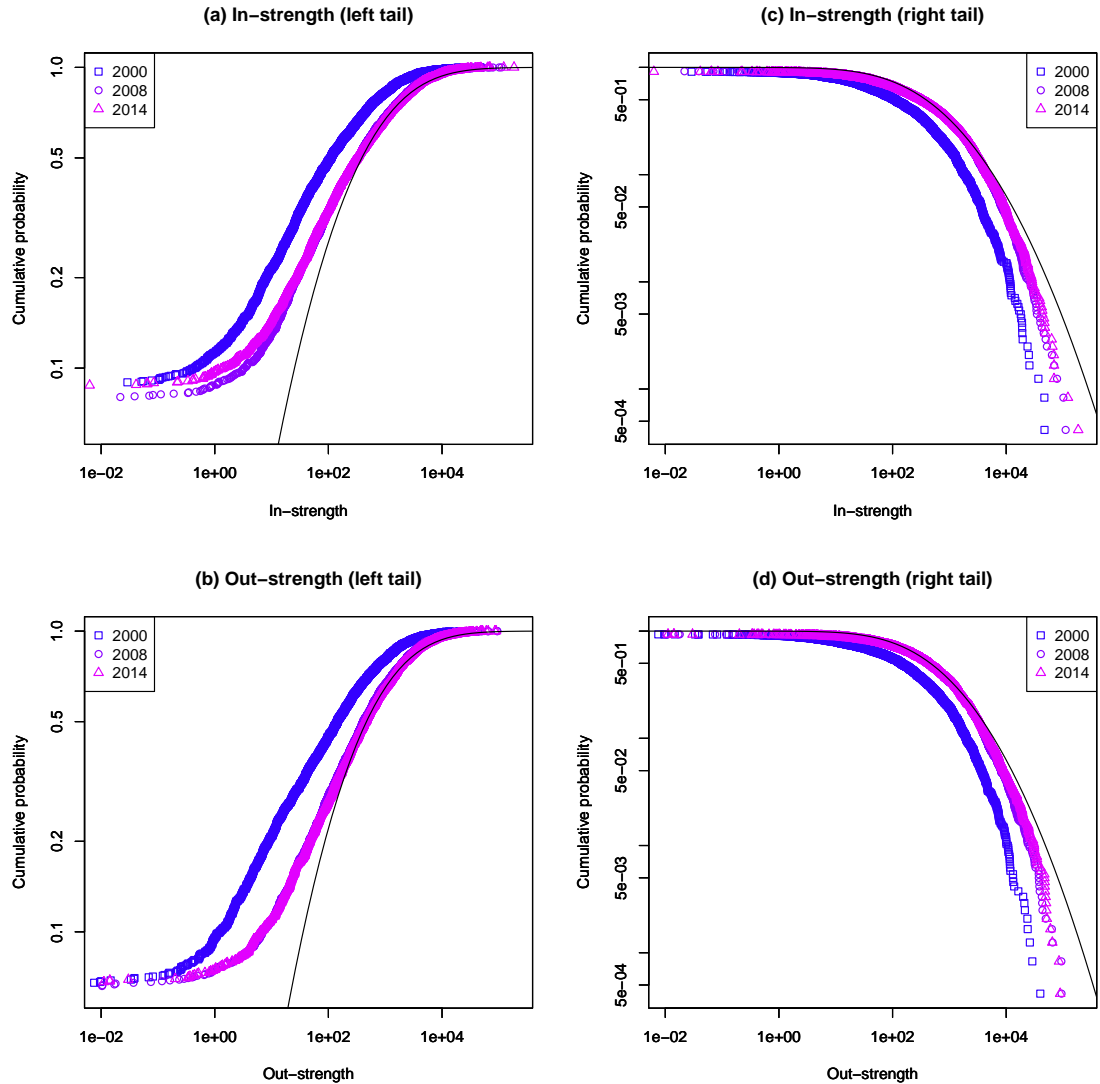}
\caption{{\bf Cumulative probability of strength of IVAN in 2000, 2008, and 2014.}
The black lines are plots of the cumulative density function of the log-normal distribution whose $\mu$ and $\sigma$ are equal to the average and the standard deviation, respectively, of the logarithm of strength distribution in 2014. Their right tails are almost fitted, but the left tails are not, especially at the bottom.}
\label{fig4}
\end{adjustwidth}
\end{figure}

Fig~\ref{fig4} shows a wide strength distribution of the IVAN's.
The distribution of Fig~\ref{fig4} partly fits a log-normal distribution, especially in the right tail. The probability density function of a log-normal distribution $p(x)$ for $x>0$ is:
\begin{equation} \label{cdf}
p(x) = \frac{1}{\sqrt{2\pi}\sigma x}\exp\biggl[-\frac{(\log x-\mu)^2}{2\sigma^2}\biggr],
\end{equation}
where $\mu$ and $\sigma$ are the mean and standard deviation of the logarithm, respectively. As seen in Fig~\ref{fig4}, the mean and standard deviation of the logarithm of strength distribution in 2014 are $\mu=5.959$ and $\sigma=2.129$ in in-strength, and $\mu=6.146$ and $\sigma=2.001$ in out-strength, respectively.

\subsection*{Potential and circular relationships in two regional communities}

\begin{figure}[!h]
\centering
\includegraphics[width=13cm,keepaspectratio]{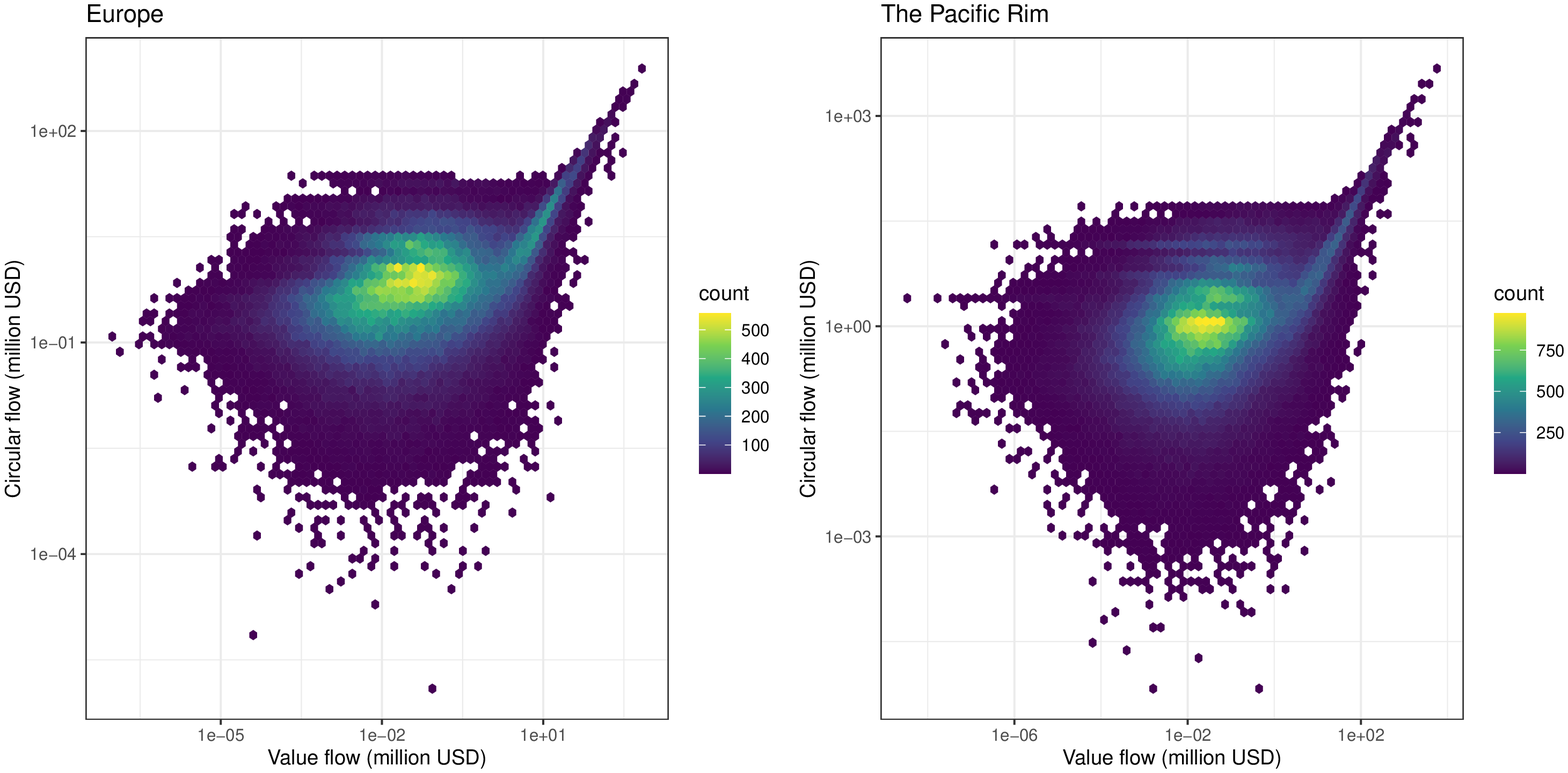}
\caption{{\bf Scatter plot of correlation between circular flow and value flow in 2000.}
The larger circular flows are in proportion to the original value flows, but the lower flows are disarranged.}
\label{fig5}
\end{figure}

Helmholtz--Hodge decomposition was applied to the two regional communities detected by Infomap.
The size of the circular flow component and the value of the IVAN link (value flow) are well correlated in the range where the value flow is greater than 100 million USD, as shown in Fig~\ref{fig5}. In the economic integration index, the amount of circular flow is divided by the GVAN weights; thus, the economic scale does not appear directly.

The potential and circular relationships in Europe and the Pacific Rim can be analyzed from the international and intersectoral perspectives by aggregating nodes into countries and sectors.

\subsubsection*{Europe}
Tables 2--7 show the breakdown of the potential and circular relationships in Europe, which is obtained from the Helmholtz--Hodge decomposition.
In terms of countries, the value potential is mainly high in Russia, Norway, and Germany and low in France, Spain, and Italy (Table~\ref{tab_hpc_e}).
From 2003 to 2012, Russia and Norway mainly have the first and second highest potential, respectively. In 2001, 2002, and 2010, with changes in the ranking of the communities in Fig~\ref{fig3}, Germany shows the highest potential.
In 2000 and 2001, 2010, and since 2013, when the European community was relatively small (Fig~\ref{fig3}), Russia had not been in Table~\ref{tab_hpc_e}. In addition, Belgium, Switzerland, and the Netherlands repeatedly appear in Table~\ref{tab_hpc_e}. Non-European countries, such as the United States, are ranked fourth from 2002 to 2005, while China is fourth in 2009 and 2010.
France is ranked the lowest-potential country except in 2002 and 2003, when Spain is ranked first, followed by Italy (Table~\ref{tab_lpc_e}). Portugal and Turkey are also ranked in half of the years. Tables~\ref{tab_hpc_e} and \ref{tab_lpc_e} show that Germany is ranked as both a high-potential and low-potential country.
\begin{table}[!ht]
\begin{adjustwidth}{-2.25in}{0in}
	\begin{center}
	\caption{{\bf Five highest-potential countries in Europe. }} \label{tab_hpc_e}
	\scalebox{0.8}[0.8]{
	\begin{tabular}{cccccccccccccccc} \hline
	Rank& 2000 & 2001 & 2002 & 2003 & 2004 & 2005 & 2006 & 2007 & 2008 & 2009 & 2010 & 2011 & 2012 & 2013 & 2014 \\ \hline
	1& NLD & DEU & DEU & RUS & RUS & RUS & RUS & RUS & RUS & RUS & DEU & RUS & RUS & POL & DEU \\
	2& DEU & BEL & RUS & NOR & NOR & NOR & NOR & NOR & NOR & NOR & GBR & NOR & NOR & ITA & RUS \\
	3& CHE & CHE & NOR & DEU & DEU & DEU & DEU & DEU & NLD & DEU & BEL & NLD & NLD & CZE & NLD \\
	4& BEL & NLD & USA & USA & USA & USA & NLD & NLD & CHE & CHN & CHN & IRL & DNK & AUT & GBR \\
	5& NOR & FIN & NLD & NLD & BEL & BEL & USA & BEL & AUT & BEL & AUT & CYP & USA & NLD & HUN \\ \hline
	\end{tabular}
	}
	 \end{center}
\end{adjustwidth}
\end{table}
\begin{table}[!ht]
\begin{adjustwidth}{-2.25in}{0in}
	\begin{center}
	\caption{{\bf Five lowest-potential countries in Europe. }} \label{tab_lpc_e}
	\scalebox{0.8}[0.8]{
	\begin{tabular}{cccccccccccccccc} \hline
	Rank& 2000 & 2001 & 2002 & 2003 & 2004 & 2005 & 2006 & 2007 & 2008 & 2009 & 2010 & 2011 & 2012 & 2013 & 2014 \\ \hline
	1& FRA & FRA & ESP & ESP & FRA & FRA & FRA & ESP & FRA & FRA & FRA & FRA & FRA & FRA & FRA \\
	2& ESP & ESP & ITA & FRA & ESP & ESP & ESP & FRA & ESP & ITA & ITA & DEU & DEU & BEL & CHE \\
	3& ITA & DNK & FRA & ITA & ITA & ITA & ITA & GBR & DEU & ESP & ESP & ITA & GBR & GBR & DNK \\
	4& PRT & PRT & POL & POL & HUN & TUR & TUR & TUR & TUR & GBR & LUX & GBR & TUR & RUS & ESP \\
	5& DNK & POL & PRT & HUN & POL & HUN & POL & ITA & POL & LUX & DNK & POL & POL & DEU & TUR \\ \hline
	\end{tabular}
	}
	 \end{center}
\end{adjustwidth}
\end{table}

Russia is ranked first in the high-potential tables (Tables~\ref{tab_hpc_e} and~\ref{tab_hps_e}) except for 2002, when sector B: mining and quarrying is ranked first. If the sector B: mining and quarrying was not considered in Table~\ref{tab_hps_e} and Russia were not included in Table~\ref{tab_hpc_e}. The other highest-potential sectors in Europe are M69: legal, accounting, and consultancy activities in 2000 and 2010; C24: manufacture of basic metals in 2001; C25: manufacture of fabricated metal products, except machinery and equipment in 2013; and G46: wholesale trade except that of motor vehicles and motorcycles in 2014. In addition, sectors N: administrative and support service activities and C20: manufacture of chemicals and chemical products are often ranked high.
In terms of low potential (Tables~\ref{tab_lpc_e} and~\ref{tab_lps_e}), sector F: construction is ranked first throughout the 15 years, followed by sectors C28: manufacture of machinery and equipment, C19: manufacture of coke and refined petroleum products, C29: manufacture of motor vehicles, which occupy the second and subsequent positions; O84: public administration and defense, and compulsory social security, which appears in the third and subsequent positions; and L68: real estate activities, which often appears in the fifth position.
\begin{table}[!ht]
\begin{adjustwidth}{-2.25in}{0in}
	\centering
	\caption{{\bf Five highest-potential sectors in Europe. }} \label{tab_hps_e}
	\scalebox{0.8}[0.8]{
	\begin{tabular}{cccccccccccccccc} \hline
	Rank& 2000 & 2001 & 2002 & 2003 & 2004 & 2005 & 2006 & 2007 & 2008 & 2009 & 2010 & 2011 & 2012 & 2013 & 2014 \\ \hline
	1& M69 & C24 & B & B & B & B & B & B & B & B & M69 & B & B & C25 & G46 \\
	2& C24 & N & N & N & N & N & N & C24 & C24 & M69 & G46 & G46 & G46 & C24 & C24 \\
	3& N & M69 & M69 & M69 & M69 & C24 & C24 & N & G46 & N & N & N & N & C22 & C25 \\
	4& C20 & C20 & C24 & C24 & C24 & M69 & M69 & G46 & M69 & G46 & C24 & C24 & M69 & G46 & C20 \\
	5& K64 & C25 & C20 & C20 & C20 & G46 & G46 & M69 & H49 & C24 & C20 & H49 & C24 & C27 & C22 \\ \hline
	\end{tabular}
	}
	\begin{flushleft} B: mining and quarrying; C20: manufacture of chemicals and chemical products; C22: manufacture of rubber and plastic products; C24: manufacture of basic metals; C25: manufacture of fabricated metal products except machinery and equipment; C27: manufacture of electrical equipment; G46: wholesale trade except trade of motor vehicles and motorcycles; H49: land transport and transport via pipelines; K64: financial service activities except insurance and pension funding; M69: legal and accounting activities; activities of head offices, and management consultancy activities; N: administrative and support service activities.
	\end{flushleft}
\end{adjustwidth}
\end{table}
\begin{table}[!ht]
\begin{adjustwidth}{-2.25in}{0in}
	\centering
	\caption{{\bf Five lowest-potential sectors in Europe. }} \label{tab_lps_e}
	\scalebox{0.8}[0.8]{
	\begin{tabular}{cccccccccccccccc} \hline
	Rank& 2000 & 2001 & 2002 & 2003 & 2004 & 2005 & 2006 & 2007 & 2008 & 2009 & 2010 & 2011 & 2012 & 2013 & 2014 \\ \hline
	1& F & F & F & F & F & F & F & F & F & F & F & F & F & F & F \\
	2& C28 & C28 & C28 & C28 & C19 & C19 & C19 & C29 & C29 & C19 & C28 & C19 & C19 & C29 & C28 \\
	3& C19 & O84 & O84 & O84 & C28 & C28 & C28 & C19 & C19 & C28 & O84 & C28 & C28 & C28 & O84 \\
	4& O84 & C31 & C19 & C19 & O84 & O84 & O84 & C28 & C28 & O84 & G47 & D35 & D35 & O84 & G47 \\
	5& C31 & L68 & L68 & L68 & L68 & R\_S & R\_S & O84 & O84 & L68 & C31 & O84 & O84 & R\_S & C33 \\ \hline
	\end{tabular}
	}
	\begin{flushleft} C19: manufacture of coke and refined petroleum products; C28: manufacture of machinery and equipment n.e.c.; C29: manufacture of motor vehicles, and trailers, and semi-trailers; C31: manufacture of furniture, and other manufacturing; C33: repair and installation of machinery and equipment; D35: electricity, gas, steam and air conditioning supply; F: construction; G47: retail trade except that of motor vehicles and motorcycles; L68: real estate activities; O84: public administration and defense, and compulsory social security; R\_S: other service activities.
	\end{flushleft}
\end{adjustwidth}
\end{table}

According to the amount of circulation (Table~\ref{tab_hlc_e}), Germany is the top country for the 15 years, followed by France, the United Kingdom, Italy, and Russia. In terms of sectoral ranking (Table~\ref{tab_hls_e}), sector F: construction is always ranked first until 2010. In 2011 and 2012, sector B: mining and quarrying, which often ranks second, is ranked first. In 2013, sector C29: manufacture of motor vehicle is first.
Russia is ranked second in the year that sector B: mining and quarrying is ranked first. Throughout the whole years, F: construction is first, followed by B: mining and quarrying, C28: manufacture of machinery and equipment, C19: manufacture of coke and refined petroleum products, and G46: wholesale trade except for that of motor vehicles and motorcycles.
\begin{table}[!ht]
\begin{adjustwidth}{-2.25in}{0in}
	\begin{center}
	\caption{{\bf Five-highest circulated countries in Europe.}} \label{tab_hlc_e}
	\scalebox{0.8}[0.8]{
	\begin{tabular}{cccccccccccccccc} \hline
	Rank& 2000 & 2001 & 2002 & 2003 & 2004 & 2005 & 2006 & 2007 & 2008 & 2009 & 2010 & 2011 & 2012 & 2013 & 2014 \\ \hline
	1& DEU & DEU & DEU & DEU & DEU & DEU & DEU & DEU & DEU & DEU & DEU & DEU & DEU & DEU & DEU \\
	2& FRA & FRA & GBR & FRA & FRA & FRA & FRA & FRA & RUS & FRA & FRA & RUS & RUS & FRA & FRA \\
	3& NLD & ITA & FRA & ITA & GBR & GBR & GBR & ITA & FRA & RUS & ITA & NOR & GBR & ITA & ITA \\
	4& ITA & ESP & ITA & GBR & ITA & ITA & RUS & RUS & ITA & GBR & ESP & FRA & NOR & ESP & POL \\
	5& ESP & GBR & ESP & ESP & NLD & RUS & ITA & GBR & NOR & ITA & NLD & GBR & FRA & POL & AUT \\ \hline
	\end{tabular}
	}
	 \end{center}
\end{adjustwidth}
\end{table}
\begin{table}[!ht]
\begin{adjustwidth}{-2.25in}{0in}
	\centering
	\caption{{\bf Five-highest circulated sectors in Europe.}} \label{tab_hls_e}
	\scalebox{0.8}[0.8]{
	\begin{tabular}{cccccccccccccccc} \hline
	Rank& 2000 & 2001 & 2002 & 2003 & 2004 & 2005 & 2006 & 2007 & 2008 & 2009 & 2010 & 2011 & 2012 & 2013 & 2014 \\ \hline
	1& F & F & F & F & F & F & F & F & B & F & F & B & B & C29 & F \\
	2& C28 & C28 & B & B & B & B & B & B & F & B & C28 & F & F & F & C28 \\
	3& G46 & G46 & N & N & C19 & C19 & C19 & C29 & C19 & C19 & G46 & C19 & C19 & C28 & G46 \\
	4& K64 & C20 & G46 & G46 & N & G46 & G46 & C19 & C29 & G46 & M69 & G46 & D35 & C25 & C25 \\
	5& C20 & C24 & C28 & C28 & G46 & C28 & C28 & G46 & G46 & D35 & K64 & D35 & G46 & C24 & C20 \\ \hline
	\end{tabular}
	}
	\begin{flushleft} B: mining and quarrying; C19: manufacture of coke and refined petroleum products; C20: manufacture of chemicals and chemical products; C24: manufacture of basic metals; C25: manufacture of fabricated metal products, except machinery and equipment; C28: manufacture of machinery and equipment n.e.c.; C29: manufacture of motor vehicles, and trailers and semi-trailers; D35: electricity, gas, steam and air conditioning supply; F: construction; G46: wholesale trade except that of motor vehicles and motorcycles; K64: financial service activities except insurance and pension funding; M69: legal and accounting activities, activities of head offices, and management consultancy activities; N: administrative and support service activities.
	\end{flushleft}
\end{adjustwidth}
\end{table}

\begin{figure}[!h]
\centering
\includegraphics[width=10cm,keepaspectratio]{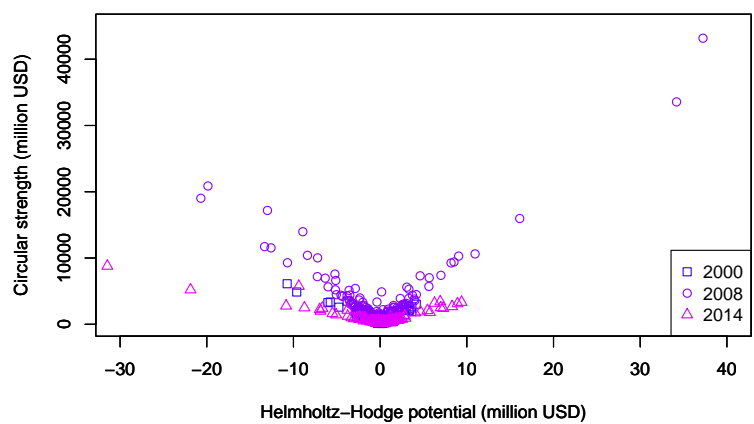}
\caption{{\bf Scatter plot of correlation between circular strength and Helmholtz--Hodge potential in Europe in 2000, 2008, and 2014.}
Each year shows a V-shaped pattern, but the inclination is changed (sharpest in 2008 and gentlest in 2014). }
\label{fig6}
\end{figure}

For both countries and sectors, those ranked higher or lower in terms of potential are also greater in terms of circular strength, which is the strength of the circular network decomposed from an IVAN.
The relationship between the circular strength and the Helmholtz--Hodge potential is shown in Fig~\ref{fig6}.
The relationship is V-shaped; in other words, the larger the absolute value of the potential, the higher the circular strength.

\subsubsection*{Pacific Rim}
The characteristics of the potential and circular flows of the IVANs in the Pacific Rim are shown in Tables 8--13.
Japan, which has the maximum potential in 2000 and 2002, has been out of the ranking since 2012; Canada shows the maximum potential from 2003 to 2007, and Australia has been the highest-potential country since then (Table~\ref{tab_hpc_p}). Indonesia, Taiwan, and India are also among the five highest-potential countries.
In terms of low potential, Mexico is the top-ranked country until 2002, the United States from 2004 to 2006, and China in 2003 and from 2007 to 2013. In addition, India, the United Kingdom, and France appear in Table~\ref{tab_lpc_p} numerous times.
\begin{table}[!ht]
\begin{adjustwidth}{-2.25in}{0in}
	\centering
	\caption{{\bf Five highest-potential countries in the Pacific Rim. }} \label{tab_hpc_p}
	\scalebox{0.8}[0.8]{
	\begin{tabular}{cccccccccccccccc} \hline
	Rank& 2000 & 2001 & 2002 & 2003 & 2004 & 2005 & 2006 & 2007 & 2008 & 2009 & 2010 & 2011 & 2012 & 2013 & 2014 \\ \hline
	1& JPN & CAN & JPN & CAN & CAN & CAN & CAN & CAN & AUS & AUS & AUS & AUS & AUS & AUS & AUS \\
	2& CAN & JPN & CAN & JPN & JPN & JPN & AUS & AUS & CAN & CAN & JPN & CAN & CAN & CAN & CAN \\
	3& AUS & AUS & AUS & AUS & AUS & AUS & JPN & JPN & IDN & JPN & CAN & IDN & IDN & TWN & TWN \\
	4& IDN & IDN & TWN & IDN & IDN & IDN & IDN & TWN & TWN & TWN & IDN & JPN & TWN & RUS & KOR \\
	5& TWN & TWN & IDN & TWN & KOR & TWN & TWN & IDN & BRA & IDN & TWN & TWN & BRA & IDN & DEU \\ \hline
	\end{tabular}
	}
\end{adjustwidth}
\end{table}
\begin{table}[!ht]
\begin{adjustwidth}{-2.25in}{0in}
	\begin{center}
	\caption{{\bf Five lowest-potential countries in the Pacific Rim. }} \label{tab_lpc_p}
	\scalebox{0.8}[0.8]{
	\begin{tabular}{cccccccccccccccc} \hline
	Rank& 2000 & 2001 & 2002 & 2003 & 2004 & 2005 & 2006 & 2007 & 2008 & 2009 & 2010 & 2011 & 2012 & 2013 & 2014 \\ \hline
	1& MEX & MEX & MEX & CHN & USA & USA & USA & CHN & CHN & CHN & CHN & CHN & CHN & CHN & USA \\
	2& CHN & CHN & USA & USA & CHN & CHN & CHN & USA & USA & MEX & MEX & IND & USA & USA & CHN \\
	3& USA & GBR & CHN & MEX & MEX & MEX & MEX & MEX & MEX & IND & GBR & MEX & IND & MEX & MEX \\
	4& GBR & IND & IND & IND & GBR & IRL & IRL & IND & KOR & GBR & IND & KOR & MEX & IDN & IND \\
	5& IND & USA & FRA & FRA & IND & IND & GBR & HUN & HUN & DEU & KOR & USA & FRA & GBR & GBR \\ \hline
	\end{tabular}
	}
	 \end{center}
\end{adjustwidth}
\end{table}

In terms of sectoral potential (Tables~\ref{tab_hps_p} and~\ref{tab_lps_p}), as in Europe, B: mining and quarrying is the highest-potential sector, and F: construction is the lowest-potential sector.
The relationship is more stable than that in Europe and remains unchanged over the 15 years.
In addition, as shown in Table~\ref{tab_hps_p}, sectors B: mining and quarrying, C20: manufacture of chemicals and chemical products, C24: manufacture of basic metals, G46: wholesale trade, K64: financial service activities, and N: administrative and support service activities are included in the high-potential sectors.
Sectors F: construction; O84: public administration and defense, and compulsory social security; C29: manufacture of motor vehicles; C10: manufacture of food products, beverages, and tobacco products; and Q: human health and social work activities are among the lowest-potential sectors until 2009 except 2001.
Since 2010, Q: human health and social work activities have risen to higher than third place.
\begin{table}[!ht]
\begin{adjustwidth}{-2.25in}{0in}
	\centering
	\caption{{\bf Five highest-potential sectors in the Pacific Rim. }} \label{tab_hps_p}
	\scalebox{0.8}[0.8]{
	\begin{tabular}{cccccccccccccccc} \hline
	Rank& 2000 & 2001 & 2002 & 2003 & 2004 & 2005 & 2006 & 2007 & 2008 & 2009 & 2010 & 2011 & 2012 & 2013 & 2014 \\ \hline
	1& B & B & B & B & B & B & B & B & B & B & B & B & B & B & B \\
	2& C20 & C20 & C20 & C20 & C20 & C24 & C24 & C24 & C24 & C24 & C20 & C24 & C24 & G46 & G46 \\
	3& C24 & C24 & C24 & C24 & C24 & C20 & C20 & G46 & G46 & C20 & G46 & C20 & G46 & C24 & C24 \\
	4& G46 & G46 & G46 & G46 & G46 & G46 & G46 & C20 & C20 & G46 & C24 & G46 & C20 & C20 & C20 \\
	5& N & N & C25 & K64 & K64 & K64 & K64 & K64 & K64 & K64 & N & N & K64 & K64 & K64 \\ \hline
	\end{tabular}
	}
	\begin{flushleft} B: mining and quarrying; C20: manufacture of chemicals and chemical products; C24: manufacture of basic metals; C25: manufacture of fabricated metal products except machinery and equipment; G46: wholesale trade except that of motor vehicles and motorcycles; K64: financial service activities except insurance and pension funding; N: administrative and support service activities.
	\end{flushleft}
\end{adjustwidth}
\end{table}
\begin{table}[!ht]
\begin{adjustwidth}{-2.25in}{0in}
	\centering
	\caption{{\bf Five lowest-potential sectors in the Pacific Rim. }} \label{tab_lps_p}
	\scalebox{0.8}[0.8]{
	\begin{tabular}{cccccccccccccccc} \hline
	Rank& 2000 & 2001 & 2002 & 2003 & 2004 & 2005 & 2006 & 2007 & 2008 & 2009 & 2010 & 2011 & 2012 & 2013 & 2014 \\ \hline
	1& F & F & F & F & F & F & F & F & F & F & F & F & F & F & F \\
	2& O84 & O84 & O84 & O84 & O84 & O84 & O84 & O84 & O84 & O84 & O84 & O84 & O84 & Q & O84 \\
	3& C29 & C29 & C29 & C29 & C29 & C29 & C29 & C29 & C29 & C29 & Q & Q & Q & O84 & Q \\
	4& C10 & Q & C10 & C10 & C10 & C10 & C10 & C10 & C10 & C10 & C29 & C29 & C29 & C29 & C29 \\
	5& Q & C10 & Q & Q & Q & Q & Q & Q & Q & Q & C10 & C10 & C10 & C10 & C10 \\ \hline
	\end{tabular}
	}
	\begin{flushleft} C10: manufacture of food products, beverages and tobacco products; C29: manufacture of motor vehicles, and trailers and semi-trailers; F: construction; O84: public administration and defense, and compulsory social security; Q: human health and social work activities.
	\end{flushleft}
\end{adjustwidth}
\end{table}

In terms of circulation (Tables~\ref{tab_hlc_p} and~\ref{tab_hls_p}), the United States is consistently in the first place, followed by Japan, Canada, and China since 2007. Since 2005, Japan has been in the fourth place, followed by Mexico.
By contrast, the sector rankings in Table~\ref{tab_hls_p} are not as consistent as those of the countries. In 2000 and 2007, C26: manufacture of computer, electronic, and optical products is ranked first; in 2008 and 2011, B: mining and quarrying is ranked first. These three sectors remain in the top three positions over the 15 years. They are followed by C29: manufacture of motor vehicles and O84: public administration and defense, and compulsory social security until 2009, and by O84: public administration and defense, and compulsory social security and Q: human health and social work activities since 2010.
\begin{table}[!ht]
\begin{adjustwidth}{-2.25in}{0in}
	\begin{center}
	\caption{{\bf Five highest-circulated countries in the Pacific Rim. }} \label{tab_hlc_p}
	\scalebox{0.8}[0.8]{
	\begin{tabular}{cccccccccccccccc} \hline
	Rank& 2000 & 2001 & 2002 & 2003 & 2004 & 2005 & 2006 & 2007 & 2008 & 2009 & 2010 & 2011 & 2012 & 2013 & 2014 \\ \hline
	1& USA & USA & USA & USA & USA & USA & USA & USA & USA & USA & USA & USA & USA & USA & USA \\
	2& JPN & CAN & CAN & CAN & CAN & CAN & CAN & CHN & CHN & CHN & CHN & CHN & CHN & CHN & CHN \\
	3& CAN & JPN & JPN & JPN & JPN & CHN & CHN & CAN & CAN & CAN & CAN & CAN & CAN & CAN & CAN \\
	4& MEX & MEX & MEX & CHN & CHN & JPN & JPN & JPN & JPN & JPN & JPN & JPN & JPN & JPN & JPN \\
	5& CHN & CHN & CHN & MEX & MEX & MEX & MEX & MEX & MEX & MEX & MEX & MEX & MEX & MEX & MEX \\ \hline
	\end{tabular}
	}
	 \end{center}
\end{adjustwidth}
\end{table}
\begin{table}[!ht]
\begin{adjustwidth}{-2.25in}{0in}
	\centering
	\caption{{\bf Five highest-circulated sectors in the Pacific Rim. }} \label{tab_hls_p}
	\scalebox{0.8}[0.8]{
	\begin{tabular}{cccccccccccccccc} \hline
	Rank& 2000 & 2001 & 2002 & 2003 & 2004 & 2005 & 2006 & 2007 & 2008 & 2009 & 2010 & 2011 & 2012 & 2013 & 2014 \\ \hline
	1& C26 & F & F & F & F & F & F & C26 & B & F & F & B & B & F & F \\
	2& F & C26 & C26 & C26 & B & B & B & F & C26 & B & B & F & F & B & B \\
	3& B & B & B & B & C26 & C26 & C26 & B & F & C26 & C26 & C26 & C26 & C26 & C26 \\
	4& C29 & O84 & C29 & O84 & O84 & C29 & O84 & O84 & O84 & O84 & O84 & O84 & O84 & O84 & O84 \\
	5& O84 & C29 & O84 & C29 & C29 & O84 & C29 & C29 & C29 & C29 & C29 & Q & Q & Q & Q \\ \hline
	\end{tabular}
	}
	\begin{flushleft} B: mining and quarrying; C26: manufacture of computer, electronic, and optical products; C29: manufacture of motor vehicles, and trailers and semi-trailers; F: construction; O84: public administration and defense, and compulsory social security; Q: human health and social work activities.
	\end{flushleft}
\end{adjustwidth}
\end{table}

\begin{figure}[!h]
\centering
\includegraphics[width=10cm,keepaspectratio]{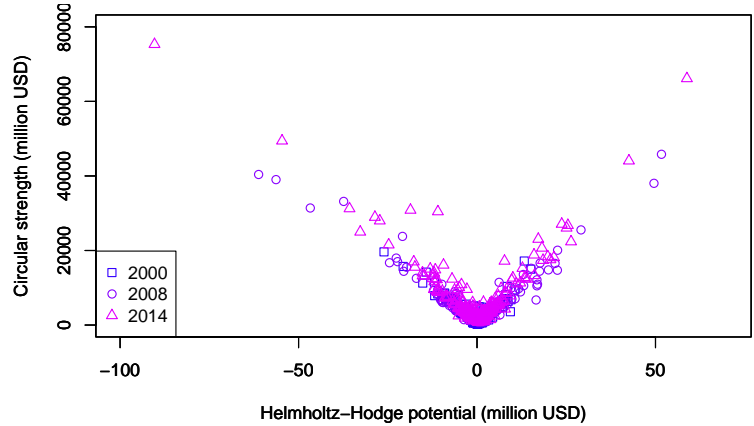}
\caption{{\bf Scatter plot of correlation between circular strength and Helmholtz-Hodge potential in the Pacific Rim in 2000, 2008, and 2014.}
The years show the similar inclinations of the V shape.}
\label{fig7}
\end{figure}

As in Europe, the high- or low-potential countries and sectors in the Pacific Rim are high in circular strength.
The relationship between the circular strength and the Helmholtz--Hodge potential is shown in Fig~\ref{fig7}.
The V-shaped relationship between them is the same as that in Europe.
However, there is a difference in the tables 2--14.
Sector C26: manufacture of computer, electronic, and optical products does not appear in Tables~\ref{tab_hps_p} and~\ref{tab_lps_p}, which indicates a large value-added circulation in the midstream of the Pacific Rim's value flow, although all high-circular-strength sectors in Europe are also in the value-added potential ranking.

\subsection*{Economic integration index}
\begin{figure}[!h]
 \centering
\includegraphics[width=13cm,keepaspectratio]{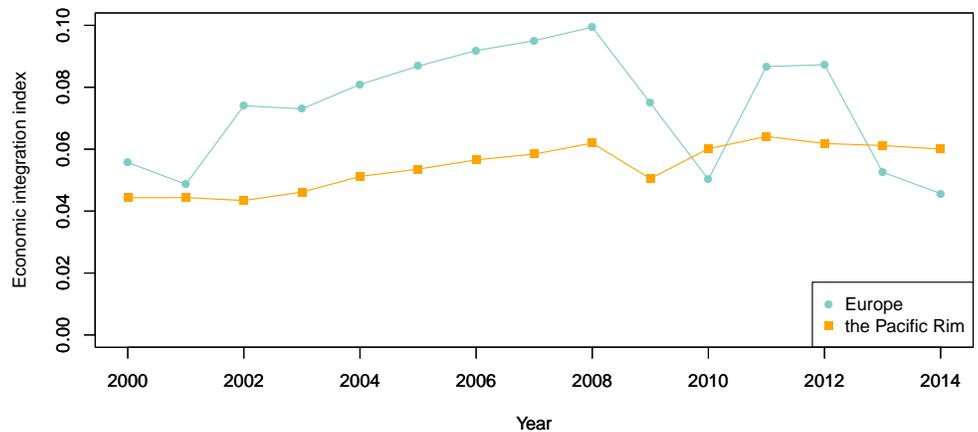}
\caption{{\bf Economic integration estimation of the two regional communities in 2000--2014.}}
\label{fig8}       
\end{figure}

The results of applying the economic integration index to the two regional communities are shown in Fig~\ref{fig8}. The Pacific Rim shows a stable and upward trend for economic integration, while Europe shows a higher but more unstable integration level. In particular, there is a large decline in the level of integration in 2009 and 2010, after the economic crisis, while the Pacific Rim is slightly affected in 2009.

\begin{figure}[!h]
\centering
\includegraphics[width=13cm,keepaspectratio]{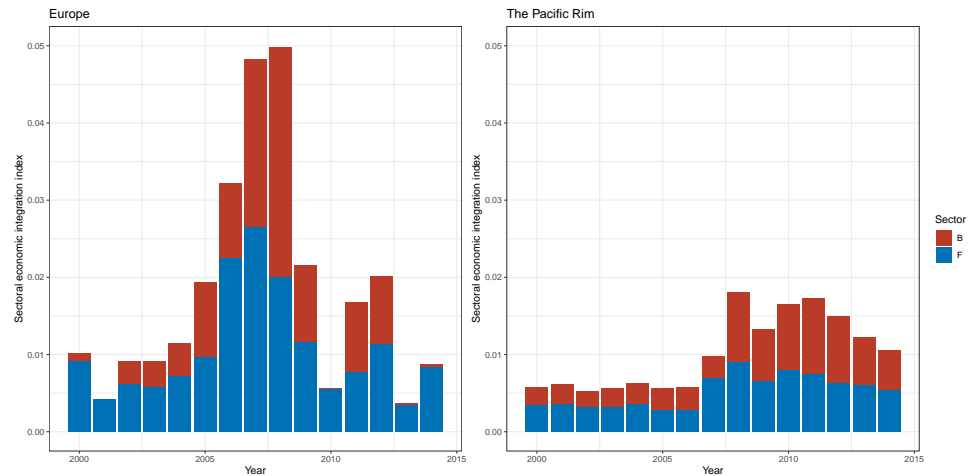}
\caption{{\bf Sectoral economic integration index of mining and quarrying, and construction.}
B: mining and quarrying, F: construction.}
\label{fig9}
\end{figure}

\begin{figure}[!h]
\centering
\includegraphics[width=13cm,keepaspectratio]{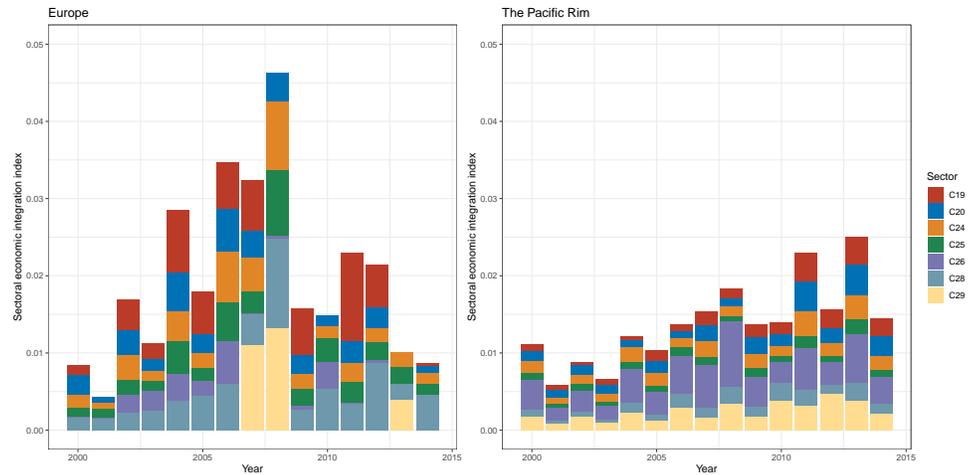}
\caption{{\bf Sectoral economic integration index of manufacture.}
C19: manufacture of coke and refined petroleum products; C20: manufacture of chemicals and chemical products; C24: manufacture of basic metals, C25: manufacture of fabricated metal products except machinery and equipment; C26: manufacture of computer, electronic, and optical products; C28: manufacture of machinery and equipment n.e.c.; C29: manufacture of motor vehicles, and trailers and semi-trailers.}
\label{fig10}
\end{figure}

\begin{figure}[!h]
\centering
\includegraphics[width=13cm,keepaspectratio]{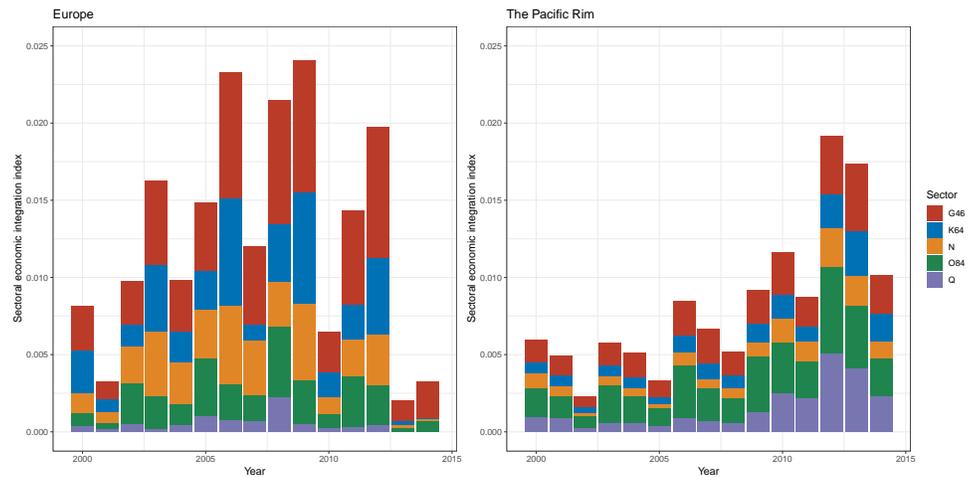}
\caption{{\bf Sectoral economic integration index of other important sectors.}
G46: wholesale trade except that of motor vehicles and motorcycles; K64: financial service activities except insurance and pension funding; N: administrative and support service activities; O84: public administration and defense, and compulsory social security; Q: human health and social work activities.}
\label{fig11}
\end{figure}

Figs~\ref{fig9}, \ref{fig10}, and \ref{fig11} show the sector-wise economic integration indices, focusing on the sectors that show high circulation in Tables~\ref{tab_hls_e} and~\ref{tab_hls_p}. As illustrated in Fig~\ref{fig9}, Europe exhibits a fivefold increase in 2007 and 2008 (just before the economic crisis) in sector B: mining and quarrying and F: construction (the top-ranking sectors in terms of potential and circular flows) compared with 2000, then a sharp decline from 2009 to 2010 (below 2000 level). The low values in 2010, 2013, and 2014 are partly due to a decline in F: construction, but the circulation within B: mining and quarrying is almost zero. By contrast, in the Pacific Rim, these values are tripled between 2007 and 2008; they decline beginning 2011, but they remain above 2007 levels in 2014.

Fig~\ref{fig10} shows the sectoral economic integration indices of the manufacturing sectors ranked in Tables~\ref{tab_hls_e} and~\ref{tab_hls_p}. The figure illustrates that the increase in the amount of circulation in the manufacturing sectors plays a major role in Europe having the largest economic integration index in 2008. Sector C29: manufacture of motor vehicles makes a significant contribution in 2007 and 2008. Sectors C24: manufacture of basic metals, C25: manufacture of fabricated metal products, and C29: manufacture of motor vehicles exhibit the highest values in 2008. The high value in 2008 can be seen from the fact that C24: manufacture of basic metals, C25: manufacture of fabricated metal products, and C29: manufacture of motor vehicles show the highest values throughout the 15 years. As for the manufacturing sectors, C19: manufacture of coke and refined petroleum products disappears from the European community at a time when the economic integration index for B: mining and quarrying is nearly zero. On the contrary, in the Pacific Rim, the manufacturing sector steadily increases its contribution to economic integration. In the years with large values in economic integration index, C19: manufacture of coke and refined petroleum products, C20: manufacture of chemicals and chemical products, and C29: manufacture of motor vehicles are large in sectoral indices.

Finally, Fig~\ref{fig11} shows the other sectors that are ranked three or more times in Tables~\ref{tab_hls_e} and~\ref{tab_hls_p}. When Europe shows a high degree of economic integration, the contributions of G46: wholesale trade, K64: financial service activities, and N: administrative and support service activities are large.
The contributions of K64: financial service activities and N: administrative and support service activities become so small after 2013, when the level of integration is low, that they are barely visible.
In the Pacific Rim, the contributions of G46: wholesale trade; O84: public administration and defense, and compulsory social security; and Q: human health and social work activities are large, although they tend to peak and decline every three years except for 2009, when the economic crisis occurred.

\section*{Discussion}
In the international trade networks used by Ikeda et al.\cite{ikeda16}, the communities are divided into industries; in original IVANs, the entire world is connected. Two buried communities---Europe and the Pacific Rim---were identified in the current paper by eliminating low flows with a threshold. However, since the IVAN included only 43 countries (excludes RoW), a more extensive international economic network, that is, one that includes other countries, may reveal more than two regional communities can.

The variability between the threshold and the community detection results is also important. In this study, we did not set the threshold as a specific weight in order to see the change over the 15 studied years. This is because the weights of IVAN links increase or decrease depending on economic growth or crisis; thus, it was inappropriate to use the same value as the threshold throughout the 15 years. In addition, as seen in Fig~\ref{fig1}, the results of the detected communities are unstable, even for close threshold values. The threshold value did not separate the state of whether a single large community or two communities appeared; rather, the trend of the distribution changed. In this study, we used a resolution of every 500 links, but more detailed research is needed on the relationship between the number of remaining links and the number of detected communities. Thus, the threshold with the most remaining links was used for this analysis.
The impact of this approach on the results is limited because the threshold was only applied for detecting concealed communities; it was not adopted for calculating the economic integration.
With the threshold value we adopted, two communities with more than 240 nodes appeared, which is around 10\% of the 2408 nodes included by the IVAN, and those communities were analyzed as regional communities. Therefore, the sector-specific European communities shown as gray nodes in Figs~\ref{fig2} and green small communities in \ref{fig3} were not treated as the regional community of Europe. In other words, the European community does not include sector-specific European communities because Europe has substantial value flows that have mainly the same type of sector, not different types of sectors.

Originally, the smile curve~\cite{mudambi} was viewed in terms of the production stage of a particular product or its relative position from a specific sector. In a broad sense, the smile curve in this study appears as a V-shaped curve when its relative position to the final goods of various sectors is considered as a whole.
Not surprisingly, Helmholtz--Hodge decomposition shows a cross-country potential flow with high (low) potential in B: mining and quarrying (f: construction), which is in the upstream (downstream) stage of production.
As seen from the circular relationship, the contributions of B: mining and quarrying and F: construction to international economic integration also indicate that the manufacture of each country depends on the resources and development demands (such as construction) of other countries.

\begin{figure}[!h]
\begin{adjustwidth}{-2.25in}{0in}
\includegraphics[width=18cm,keepaspectratio]{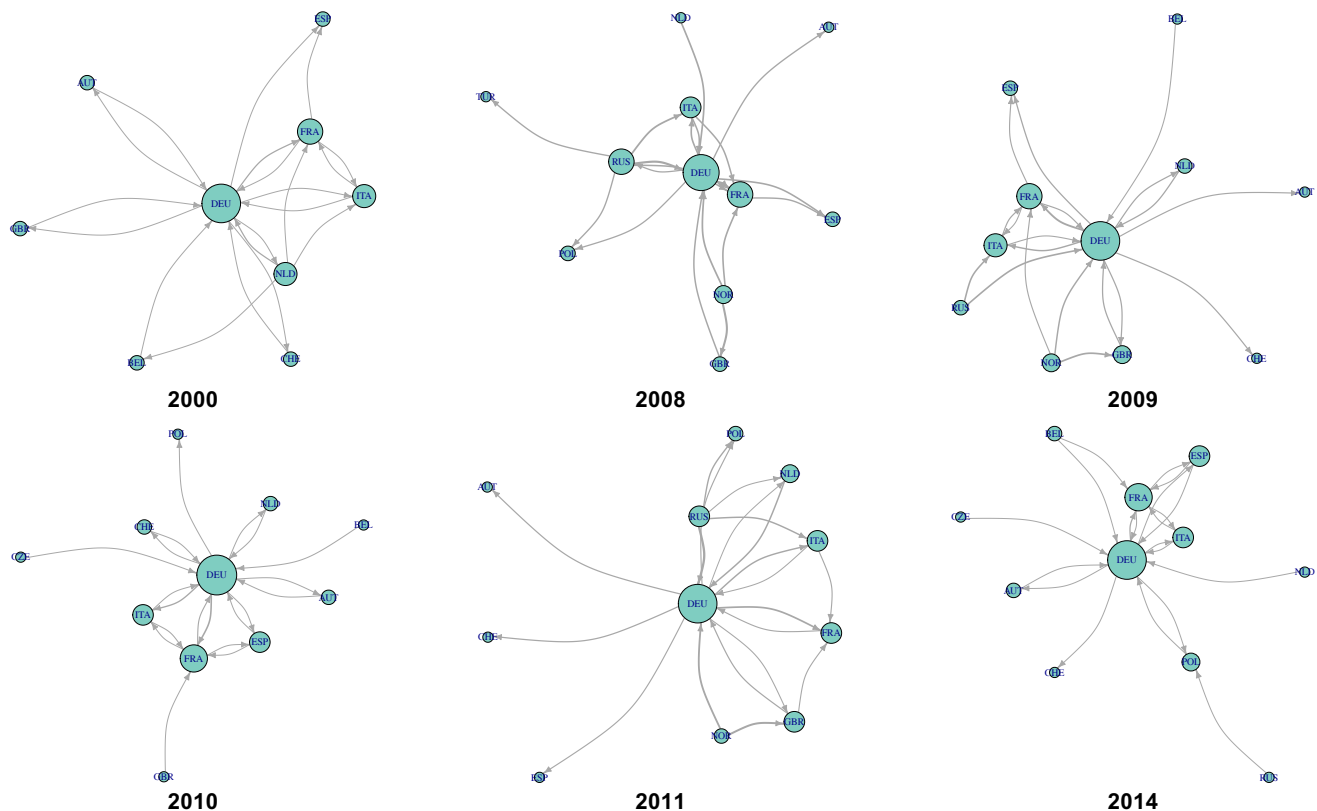}
\caption{{\bf International value-added circulation of Europe in 2000, 2008--2011, and 2014.}
These graphs illustrate the top 20 links. Their node sizes and link widths are in proportion to the square root of the degree and the amount of circular flow, respectively. \nameref{S1_Table} shows a detailed list of the sectors.}
\label{fig12}
\end{adjustwidth}
\end{figure}

\begin{figure}[!h]
\begin{adjustwidth}{-2.25in}{0in}
\includegraphics[width=18cm,keepaspectratio]{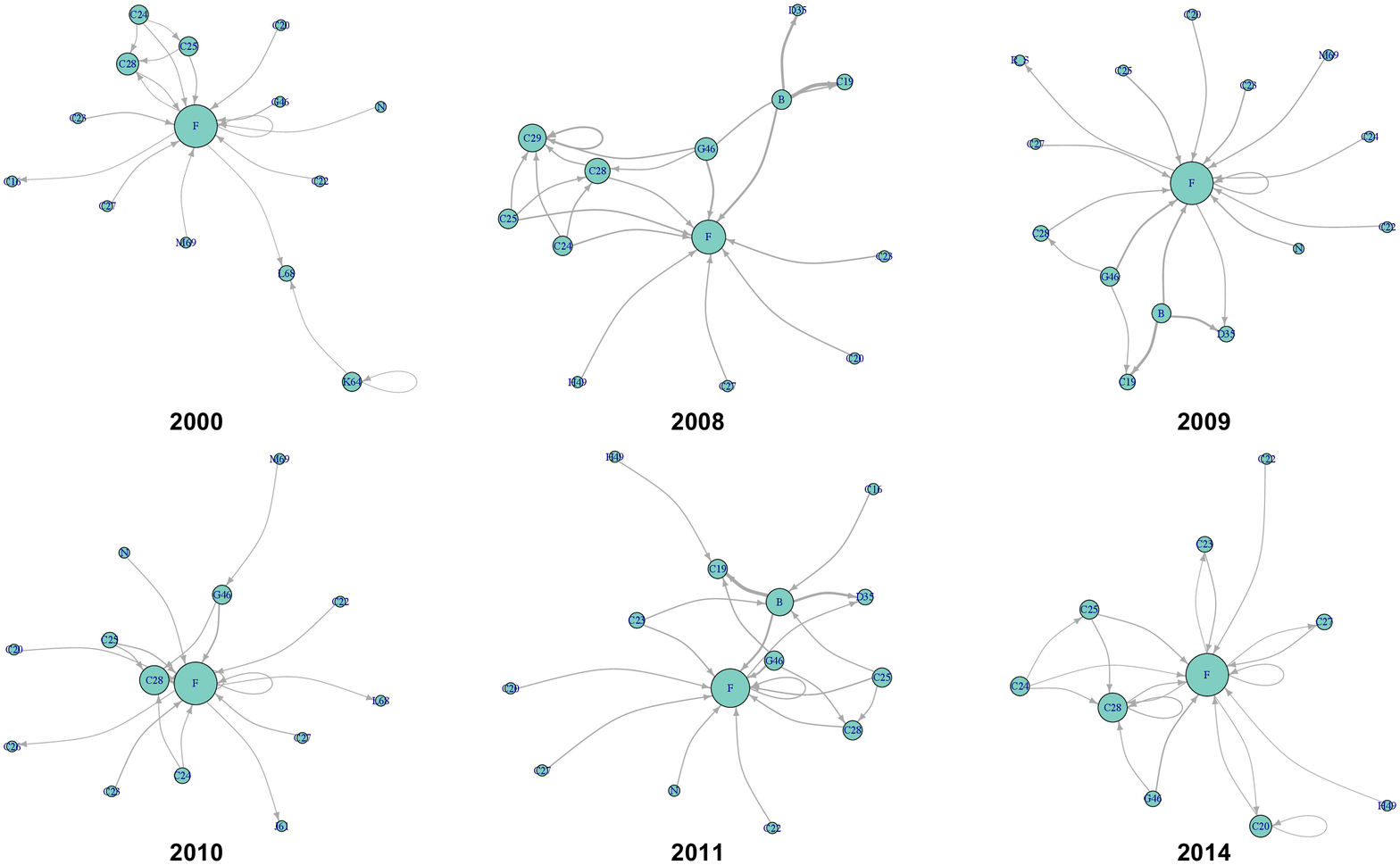}
\caption{{\bf Intersectoral value-added circulation of Europe in 2000, 2008--2011, and 2014.}
These graphs illustrate the top 20 links. Their node sizes and link widths are in proportion to the square root of the degree and the amount of circular flow, respectively. \nameref{S2_Table} shows a detailed list of the sectors.}
\label{fig13}
\end{adjustwidth}
\end{figure}

To examine the implications of the economic integration index, we focused on the period of 2008--2011, which has exhibited substantial changes in economic integration (Fig~\ref{fig8}); we also analyzed 2000 and 2014, which are the first and last years of the analyzed WIOD.
Fig~\ref{fig12} shows the changes in the international and intersectoral relationships of circular flow in Europe. The circulation in 2008, the year of the highest integration index in Europe for the 15 years, is higher between Germany, France, Italy, and Russia compared with that in 2000. However, in 2011 and 2014, which has the lowest economic integration, Russia is absent in Europe, as shown in Tables~\ref{tab_hpc_e} and \ref{tab_hlc_e}. From a sectoral point of view, Fig~\ref{fig13} shows sector B: mining and quarrying as a hub of sectors F: construction; C19: manufacture of coke and refined petroleum products; and D35: electricity, gas, steam, and air conditioning supply. Therefore, the role of Russia and sector B: mining and quarrying is important for the high economic integration index in Europe.

\begin{figure}[!h]
\begin{adjustwidth}{-2.25in}{0in}
\includegraphics[width=18cm,keepaspectratio]{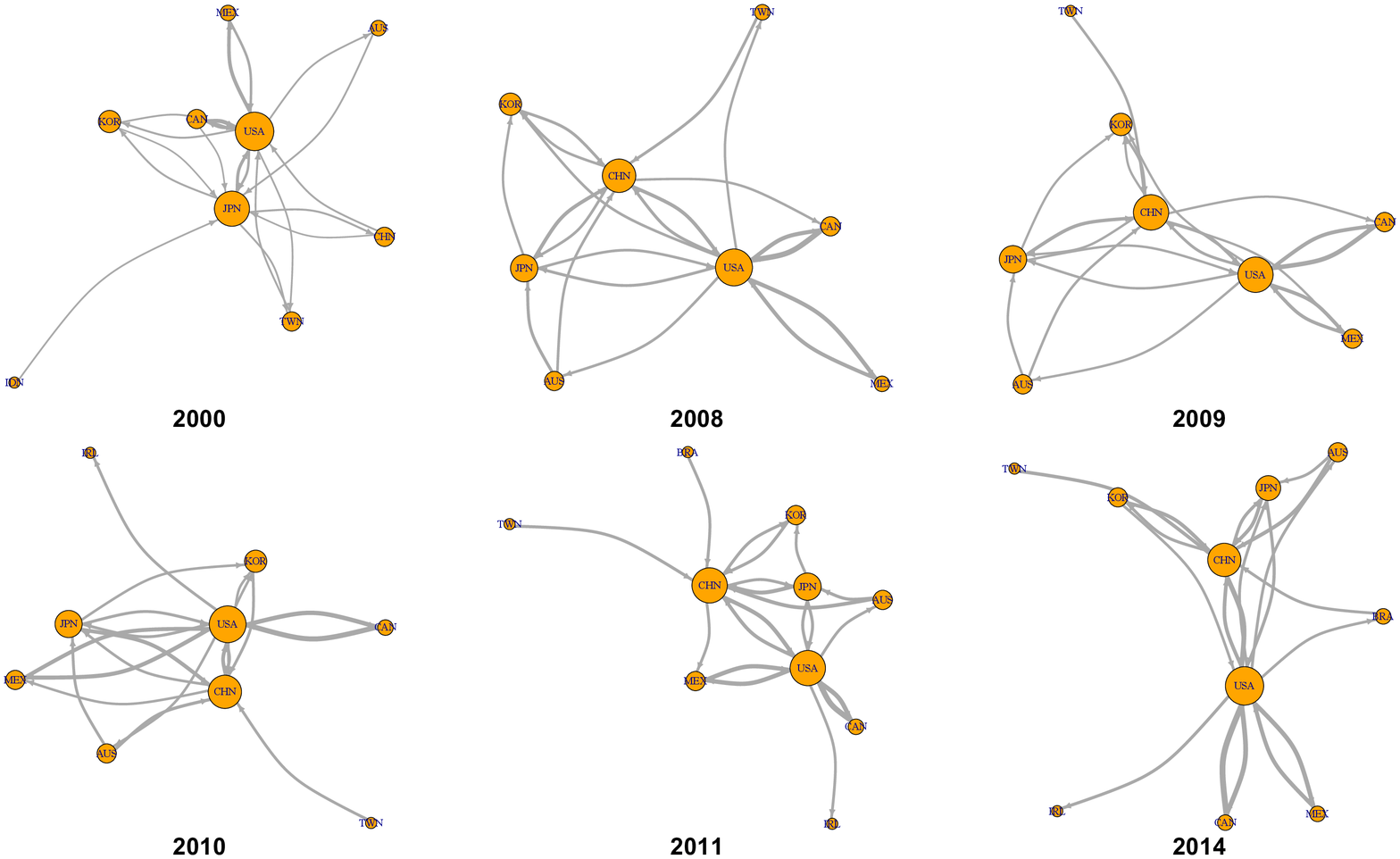}
\caption{{\bf International value-added circulation of the Pacific Rim in 2000, 2008--2011, and 2014.}
These graphs illustrate the top 20 links. Their node sizes and link widths are in proportion to the square root of the degree and the amount of circular flow, respectively. \nameref{S1_Table} shows a detailed list of the sectors.}
\label{fig14}
\end{adjustwidth}
\end{figure}

\begin{figure}[!h]
\begin{adjustwidth}{-2.25in}{0in}
\includegraphics[width=18cm,keepaspectratio]{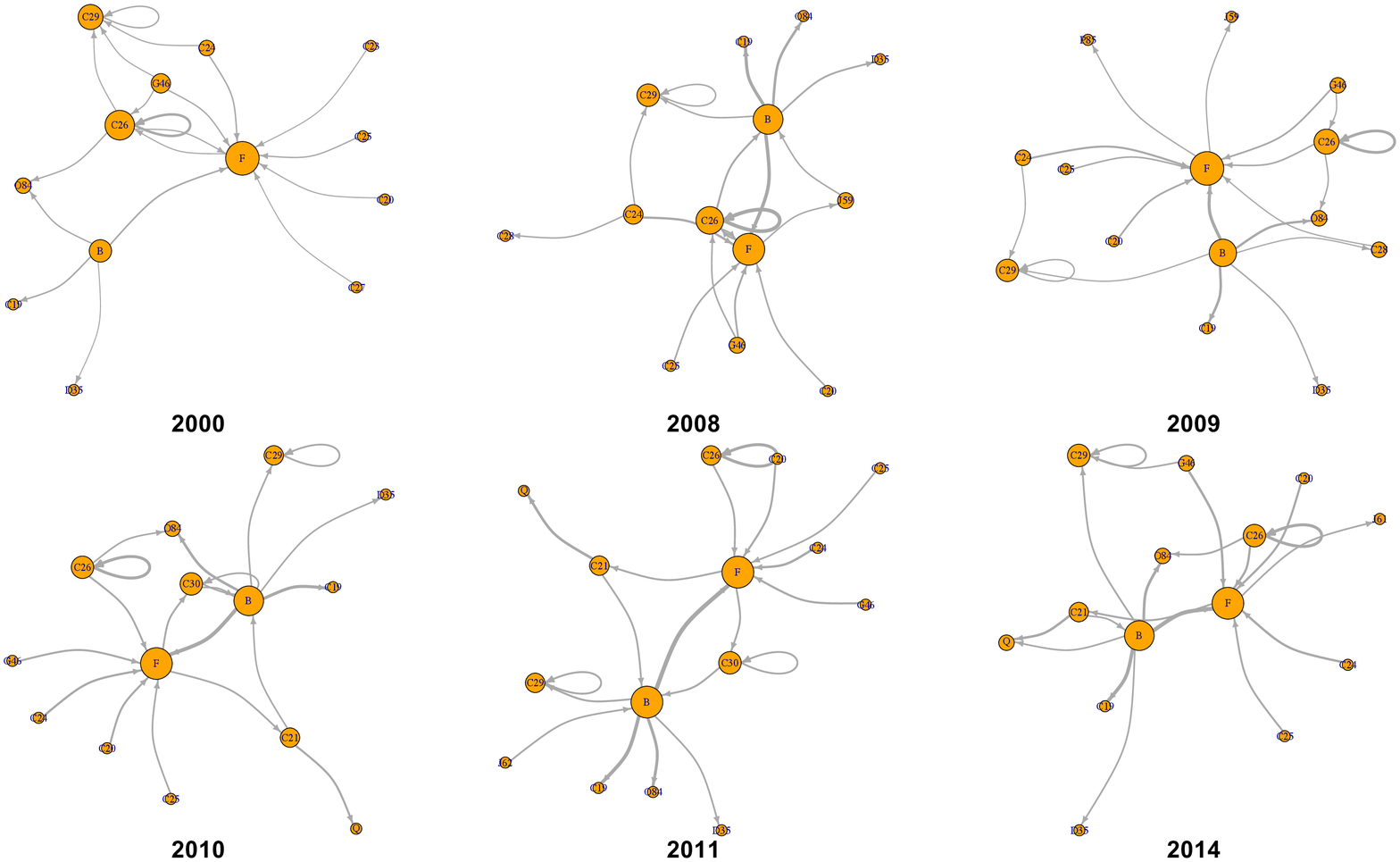}
\caption{{\bf Intersectoral value-added circulation of the Pacific Rim in 2000, 2008--2011, and 2014.}
These graphs illustrate the top 20 links. Their node sizes and link widths are in proportion to the square root of the degree and the amount of circular flow, respectively. \nameref{S2_Table} shows a detailed list of the sectors.}
\label{fig15}
\end{adjustwidth}
\end{figure}

Figs~\ref{fig14} and~\ref{fig15} show the changes in international and intersectoral relationships in the Pacific Rim. According to Fig~\ref{fig14}, the center of the value-added circulation in the Pacific Rim changed from the United States and Japan to the United States and China. This also means that the Pacific Rim, as a regional community of IVANs, was detected stably because of the strong value-added circulation around the United States and China despite the economic crisis around 2009. From the sectoral perspective, Fig~\ref{fig15} shows that there are three crucial points in the circular structure of the Pacific Rim; these are the strongest stable circulations from B: mining and quarrying to F: construction, those within C26: manufacture of computer, electronic, and optical products, and those within C29: manufacture of motor vehicles, and trailers and semi-trailers for the years 2000--2014.

In Europe, the circulation between sectors B: mining and quarrying and F: construction is unstable because sector B mainly occurs in Russia, which is not an EU member country. Therefore, the extent of economic integration tends to be unstable.
By contrast, the Pacific Rim has mineral resources.
It also has neutral- and low-potential sectors with high circulation, namely, manufacture of high-tech and motor vehicles, respectively, which produces the high-value-added products as parts of GVCs.

These results suggest that the stable growth of economic integration in the Pacific Rim has been driven by the relationship of the value-added circulation of these resources and international division in the manufacture of high value-added products.
Furthermore, in the Pacific Rim, the international division of labor is advancing, and value-added circulation is occurring across countries.
This can be explained by the fact that, the free movement of labor in the EU has led to the specialization of sectors; in the Pacific Rim, the movement of labor is limited, and the international division of labor has led to the free movement of goods.

\section*{Conclusion}
The purpose of this study is to clarify how international economic integration is occurring from the perspective of trade in value-added.
For this purpose, we used the WIOD from 2016 to construct and analyze IVANs, which show the international relationship of sector-wise trade in value-added. First, the scope of economic integration was identified by Infomap, a community detection method that uses network flows. With threshold setting in the IVANs, regional communities in Europe and the Pacific Rim were detected throughout the 15 studied years. These two communities have not been found in studies that analyzed international trade networks constructed from the WIOD.
To analyze how value flows within these two regions, we used Helmholtz--Hodge decomposition to extract the potential and circular relationships and clarified the annual changes in the roles played by the countries and sectors within these regions. In addition, we defined an economic integration index using the circular flow and applied it to both regions. We found that the level of economic integration in Europe, which had been increasing until 2008, dropped sharply after the economic crisis in 2009 to a level lower than that of the Pacific Rim in 2010, recovered in 2011, and dropped again after 2012 to a level below that of the Pacific Rim in 2013 and 2014. While the level of economic integration in Europe has been unstable, that in the Pacific Rim has been on a stable upward trend.

Moreover, the sectoral economic integration index provides a background to the changes in the extent of economic integration of these two regions. In Europe, the extent of economic integration declined in 2009, 2010, and 2013 due to the decrease of intra-European value-added circulation through the industries of mining, construction, petroleum, metal, machinery manufacture, wholesale, financial services, and management services, which showed a large amount of the circular flow of IVAN.
This unstable extent of economic integration in Europe can be attributed to the fact that the free movement of labor within the EU has resulted in the connection of value-added circulation in specific sectors rather than the region as a whole. Meanwhile, in the Pacific Rim, economic integration is progressing slowly but steadily against the international vertical division of labor in sectors C26: manufacture of computer, electronic, and optical products, and C29: manufacture of motor vehicles. In the Pacific Rim, the international division of labor is believed to be developing stably because of the limited movement of labor.

\section*{Supporting information}
%
\paragraph*{S1 Table.}
\label{S1_Table}
{\bf List of countries and regional classification.} The table is in Appendix as Table~\ref{S1_Tab}.

\paragraph*{S2 Table.}
\label{S2_Table}
{\bf List of Sectors.}  The table is in Appendix as Table~\ref{S2_Tab}.

\section*{Acknowledgments}
This work was supported by JSPS KAKENHI Grant Numbers JP17KT0034.

%
%
%

\appendix
\section*{Appendix}
%

\begin{table}[!ht]
	\begin{center}
	\caption{{\bf Country list and regional classification.} } \label{S1_Tab}
	\scalebox{0.9}[0.9]{
	\begin{tabular}{ccc|ccc} \hline
	ISO code& Short name & Region & ISO code& Short name & Region \\ \hline
	AUS& Australia  & Pacific Rim & IRL& Ireland & Europe \\
	AUT& Austria  & Europe & ITA& Italy & Europe \\
	BEL& Belgium  & Europe & JPN& Japan & Pacific Rim \\
	BGR& Bulgaria  & Europe & KOR& Korea & Pacific Rim \\
	BRA& Brazil  & Pacific Rim & LTU& Lithuania & Europe \\
	CAN& Canada  & Pacific Rims & LUX& Luxembourg & Europe \\
	CHE& Switzerland  & Europe & LVA& Latvia & Europe \\
	CHN& China  & Pacific Rim & MEX& Mexico & Pacific Rim \\
	CYP& Cyprus  & Europe &MLT& Malta & Europe \\
	CZE& Czechia  & Europe &NLD& Netherlands & Europe \\
	DEU& Germany  & Europe &NOR& Norway & Europe \\
	DNK& Denmark  & Europe & POL& Poland & Europe \\
	ESP& Spain  & Europe & PRT& Portugal & Europe \\
	EST& Estonia  & Europe & ROU& Romania & Europe \\
	FIN& Finland  & Europe & RUS& Russia & Pacific Rim \\
	FRA& France & Europe & SVK& Slovakia & Europe \\
	GBR& United Kingdom & Europe & SVN& Slovenia & Europe \\
	GRC& Greece & Europe & SWE& Sweden & Europe \\
	HRV& Croatia  & Europe & TUR& Turkey & Europe \\
	HUN& Hungary & Europe & TWN& Taiwan & Pacific Rim \\
	IDN & Indonesia & Pacific Rim & USA& America & Pacific Rim \\
	IND& India & Pacific Rim & & & \\ \hline
	\end{tabular}
	 }
	 \end{center}
\end{table}

\begin{table}[!ht]
	\begin{center}
	\caption{{\bf S2 Table: List of sectors.}} \label{S2_Table}
	\scalebox{0.9}[0.9]{
	\small
	\begin{tabular}{cl} \hline
	Code & Description \\ \hline
	A01 & Crop and animal production, hunting and related service activities \\
	A02 & Forestry and logging \\
	A03 & Fishing and aquaculture \\
	B & Mining and quarrying \\
	C10 & Manufacture of food products, beverages and tobacco products \\
	C13 & Manufacture of textiles, wearing apparel and leather products \\
	\multirow{2}{*}{C16} & Manufacture of wood and of products of wood and cork except furniture, \\
	 & and manufacture of articles of straw and plaiting materials \\
	C17 & Manufacture of paper and paper products \\
	C18 & Printing and reproduction of recorded media \\
	C19 & Manufacture of coke and refined petroleum products \\
	C20 & Manufacture of chemicals and chemical products \\
	C21 & Manufacture of basic pharmaceutical products and pharmaceutical preparations \\
	C22 & Manufacture of rubber and plastic products \\
	C23 & Manufacture of other non-metallic mineral products \\
	C24 & Manufacture of basic metals \\
	C25 & Manufacture of fabricated metal products except machinery and equipment \\
	C26 & Manufacture of computer, electronic, and optical products \\
	C27 & Manufacture of electrical equipment \\
	C28 & Manufacture of machinery and equipment n.e.c. \\
	C29 & Manufacture of motor vehicles, trailers and semi-trailers \\
	C30 & Manufacture of other transport equipment \\
	C31 & Manufacture of furniture, and other manufacturing \\
	C33 & Repair and installation of machinery and equipment \\
	D35 & Electricity, gas, steam and air conditioning supply \\
	E36 & Water collection, treatment and supply \\
	\multirow{2}{*}{E37} & Sewerage, waste collection, treatment and disposal activities, and materials recovery, \\
	 & and remediation activities and other waste management services \\
	F & Construction \\
	G45 & Wholesale and retail trade and repair of motor vehicles and motorcycles \\
	G46 & Wholesale trade except that of motor vehicles and motorcycles \\
	G47 & Retail trade except that of motor vehicles and motorcycles \\
	H49 & Land transport and transport via pipelines \\
	H50 & Water transport \\
	H51 & Air transport \\
	H52 & Warehousing and support activities for transportation \\
	H53 & Postal and courier activities \\
	I & Accommodation and food service activities \\
	J58 & Publishing activities \\
	\multirow{2}{*}{J59} & Motion picture, video and television programme production, sound recording and \\
	 & music publishing activities; programming and broadcasting activities \\
	J61 & Telecommunications \\
	J62 & Computer programming, consultancy and related activities; information service activities \\
	K64 & Financial service activities, except insurance and pension funding \\
	K65 & Insurance, reinsurance and pension funding except compulsory social security \\
	K66 & Activities auxiliary to financial services and insurance activities \\
	L68 & Real estate activities \\
	\multirow{2}{*}{M69} & Legal and accounting activities, and activities of head offices, and management \\
	 & consultancy activities \\
	M71 & Architectural and engineering activities, and technical testing and analysis \\
	M72 & Scientific research and development \\
	M73 & Advertising and market research \\
	M74 & Other professional, scientific and technical activities, and veterinary activities \\
	N & Administrative and support service activities \\
	O84 & Public administration and defense, and compulsory social security \\
	P85 & Education \\
	Q & Human health and social work activities \\
	R\_S & Other service activities \\
	\multirow{2}{*}{T} & Activities of households as employers, and undifferentiated goods and services \\
	 & producing activities of households for own use \\
	U & Activities of extraterritorial organizations and bodies \\ \hline
	\end{tabular}
	 }
	 \end{center}
\end{table}

\end{document}